\documentclass[twocolumn]{aastex631}
\usepackage{amsmath}

\shorttitle{Maximal GRB Jet Energy}
\shortauthors{Wu et al.}

\begin{document}

\title{Maximal Jet Energy of Gamma-Ray Bursts through the Blandford–Znajek Mechanism}

\author[0000-0002-0196-9169]{Zhao-Feng Wu}
\affiliation{Department of Physics and Astronomy, Purdue University, 525 Northwestern Avenue, West Lafayette, IN 47907, USA}
\correspondingauthor{Zhao-Feng Wu}
\email{wu2177@purdue.edu}
\author[0000-0003-4583-6023]{Michail Damoulakis}
\affiliation{Department of Physics and Astronomy, Purdue University, 525 Northwestern Avenue, West Lafayette, IN 47907, USA}
\author[0000-0001-7833-1043]{Paz Beniamini}
\affiliation{Department of Natural Sciences, The Open University of Israel, P.O. Box 808, Ra’anana 4353701, Israel}
\affiliation{Astrophysics Research Center of the Open University (ARCO), The Open University of Israel, P.O. Box 808, Ra’anana 4353701, Israel}
\affiliation{Department of Physics, The George Washington University, 725 21st Street NW, Washington, DC 20052, USA}
\author[0000-0003-1503-2446]{Dimitrios Giannios}
\affiliation{Department of Physics and Astronomy, Purdue University, 525 Northwestern Avenue, West Lafayette, IN 47907, USA}

\begin{abstract}

Gamma-ray bursts (GRBs) are among the most energetic events in the universe, driven by relativistic jets launched from black holes (BHs) formed during the collapse of massive stars or after the merger of two neutron stars (NSs). The jet power depends on the BH spin and the magnetic flux accreted onto it. In the standard thin disk model, jet power is limited by insufficient magnetic flux, even when the spin approaches maximum possible value. In contrast, the magnetically arrested disk (MAD) state limits jet energy by extracting significant angular momentum, braking BH rotation. We propose a unified model incorporating both standard thin disk and MAD states, identifying a universal curve for jet power per accretion rate as a function of the magnetic flux ratio, $\Delta_\mathrm{eq} = (\Phi_\mathrm{BH}/\Phi_\mathrm{MAD})_\mathrm{eq}$, at spin equilibrium. For long GRBs (lGRBs), the model predicts a maximum jet energy of $\sim$1.5\% of the accretion energy, occurring at $\Delta_\mathrm{eq} \sim 0.4$ where the BH equilibrium spin is $a \sim 0.5$. Both long and short GRBs are unlikely to be produced by a MAD: for short GRBs (sGRBs), this requires an accreted mass orders of magnitude smaller than that available, while for lGRBs, the narrow progenitor mass distribution challenges the ability to produce the observed broad distribution of jet energies. This framework provides a consistent explanation for both standard and luminous GRBs, emphasizing the critical role of magnetic flux. Both long and short GRBs require magnetic flux distributions that peak around $10^{27}\,\mathrm{G\,cm}^2$.

\end{abstract}


\section{Introduction} \label{intro}
Gamma-ray bursts (GRBs) are among the most energetic events in the universe. Long duration GRBs (lGRBs) are connected to the core collapse of massive stars, as demonstrated by their association with supernovae \citep{2003ApJ...591L..17S, 2017AdAst2017E...5C, 2022ApJ...938...41D}. These supernovae, typically Type Ic, indicate Wolf-Rayet (WR) stars of spectral type WO/C as progenitors \citep{2006ARA&A..44..507W}. On the other hand, short duration GRBs (sGRBs) are linked to the mergers of neutron stars (NSs; \citealp{1989Natur.340..126E}), a connection confirmed by gravitational wave detections by LIGO \citep{2017PhRvL.119p1101A, 2017ApJ...848L..13A} and the subsequent kilonova \citep[see, e.g.,][]{2017ApJ...848L..17C}.

Despite progress in understanding GRB progenitors, the nature of their central engine remains uncertain. GRBs may be powered by accretion onto a black hole (BH; \citealp{1993ApJ...405..273W,2003astro.ph.12347L, 2015ApJS..218...12L}) or by the rapid rotation of a magnetar \citep{1992Natur.357..472U,2011MNRAS.413.2031M}.  Here, we focus on the Blandford–Znajek (BZ; \citealp{1977MNRAS.179..433B}) process, in which the jet is launched by magnetic fields extracting the rotational energy of a BH. The jet power in the BZ mechanism scales with the square of the BH spin and the magnetic flux, which is sustained by strong accretion \citep[see, e.g.,][]{2024arXiv240716745G}. When the magnetic field is sufficiently strong, the system reaches the magnetically arrested disk (MAD) state, where the magnetic flux is regulated by the accretion rate \citep{2011MNRAS.418L..79T}. While many studies use MAD to model GRBs \citep{2022A&A...668A..66J, 2022ApJ...935..176J, 2024ApJ...961..212J}, the jet efficiency becomes too high when the BH spin is not significantly low \citep{2023ApJ...952L..32G,2024ApJ...961..212J}. Typical inferred jet energies for short and long GRBs require accreted masses of $M_\mathrm{acc} \sim 10^{-4} M_\odot$ and $M_\mathrm{acc} \sim 10^{-2} M_\odot$, respectively, assuming a BH spin of $a = 0.5$ \citep{2024ApJ...960...82L}. These accreted masses are extremely small compared to the mass available for accretion in sGRBs ($\sim 0.1 M_\odot$) and lGRBs ($\sim 10 M_\odot$).

This tension of the energetics predicted by the MAD model and observations is mitigated when considering the evolution of BH spin during the GRB, determined by hydrodynamic torques from accreting matter and electromagnetic torques from the jet \citep{2024ApJ...960...82L}. In long GRBs, sufficient accretion allows the BH to reach an equilibrium spin determined by the balance of these torques. Under MAD conditions, the jet efficiently removes angular momentum, reducing spin and limiting jet efficiency before the jet emerges from the stellar envelope \citep{2024ApJ...961..212J}. This helps resolve the issue of excessive efficiency. For short GRBs, however, BHs formed from NS mergers have high initial spins \citep{Bernuzzi2014, 2020MNRAS.497.1488B}, but the available accretion mass is too small to significantly reduce the spin, leading to overly efficient jets. This indicates the need for a more comprehensive model beyond MAD.

Models involving magnetars and neutrino annihilation for powering a GRB have upper limits on the energy that can be produced \citep{2014MNRAS.443...67M, 2014MNRAS.445L...1L,2017MNRAS.472.3058B}. For example, even accounting for the additional energy source provided by accretion, magnetar central engines correspond to a maximum GRB jet energy of $\lesssim 3 \times 10^{52}$\,erg \citep{2018ApJ...857...95M}. The BZ mechanism, however, does not have a clear upper limit. In a weak magnetic field, the BH can spin up rapidly to near maximum, but jet power is constrained by insufficient magnetic flux \citep{2015MNRAS.447..327T}. In the MAD state, the system reaches maximal magnetic flux for a given accretion rate, but the jet significantly brakes the BH, reducing its spin to $a_\mathrm{eq} \sim 0.06$, which limits the jet power \citep{2024ApJ...961..212J}. Although strong magnetic flux and high spin are both essential for producing powerful jets, they work against each other, preventing MAD from generating the most powerful jets at a given accretion rate. This interplay leads to a transition point where jet power is maximized at moderate levels of both spin and magnetic flux, marking the shift from a standard thin disk to the MAD state. Our goal is to model this transition, enabling predictions of maximum jet power (or energy) for a given accretion rate (or accretion mass).

In this \textit{Letter}, we outline jet power in Section~\ref{jet_power} and present a model for BH spin and mass evolution in Section~\ref{BH_evo}, offering a semi-analytical approach that transitions from the standard thin disk to MAD. In Section~\ref{lGRBs}, we apply this model to lGRBs and compare results with observations, while Section~\ref{sGRBs} covers its application to sGRBs. Key assumptions and results are discussed in Section~\ref{discussion}, and conclusions are summarized in Section~\ref{conclude}. Gaussian-cgs units are used throughout.

\section{Jet Power} \label{jet_power}
In the BZ framework, GRB jets are driven by the extraction of rotational energy from a central BH, facilitated by large-scale magnetic flux threading the BH. This study focuses on determining the maximum jet power achievable during the collapse of massive stars via the BZ process:
\begin{equation} \label{L_BZ}
L_{\mathrm{BZ}} \approx \frac{\kappa f}{4 \pi c} \Phi_{\mathrm{BH}}^2 \Omega_{\mathrm{H}}^2,
\end{equation}
where \(\kappa \approx 0.05\) for a parabolic magnetic field geometry, $a$ is the BH spin, \(\Omega_\mathrm{H} = ac/(2r_\mathrm{H})\) is the angular frequency of the BH horizon, and \(\Phi_{\mathrm{BH}}\) denotes the magnetic flux threading the BH. Here $r_{\mathrm{H}}=r_{\mathrm{g}}\left[1+\left(1-a^2\right)^{1 / 2}\right]$ is the radius of BH horizon and $r_{\mathrm{g}}$ is the gravitational radius. We use the sixth-order expression `BZ6' from \cite{2010ApJ...711...50T}, which accurately models jet power across all spin values. The correction factor is given by $f=1+0.35 \omega_{\mathrm{H}}^2- 0.58 \omega_{\mathrm{H}}^4$, where $\omega_{\mathrm{H}}=a /\left[1+\left(1-a^2\right)^{1 / 2}\right]$ is the dimensionless rotational frequency of the BH event horizon. 

As the collapsing star's mass accretes onto the BH, the magnetic flux carried by the infalling matter also accumulates around the BH. This process leads to the formation of a MAD when the magnetic pressure becomes comparable to the gravitational pull of the accreting material. In the MAD regime, the jet power is no longer directly dependent on the magnetic flux, but is instead governed by the accretion rate:
\begin{equation} \label{L_MAD}
L_{\mathrm{MAD}} = \eta_{\mathrm{EM}}(a) \dot{m} c^2,
\end{equation}
where \(\eta_{\mathrm{EM}}(a)\) is the jet launching efficiency, derived from non-radiative 3D general relativistic magnetohydrodynamic (GRMHD) global simulations of MADs by \cite{2024ApJ...960...82L}. The jet efficiency \(\eta_{\mathrm{EM}}\) is expressed as:
\begin{equation} \label{eta_EM}
\eta_{\mathrm{EM}} \times 100 = 
\left\{
\begin{aligned}
-19.8 a^4 + 48.9 a^2, & \text{  if } a \leq 0, \\
106.3 a^4 + 39.5 a^2, & \text{  if } a \geq 0.
\end{aligned}
\right.
\end{equation}
The transition to the MAD state occurs when the jet power approaches the MAD value for a given spin. Then the flux onto the BH becomes dynamically important and diffuses outwards. This is the maximally allowed flux maintained by a given mass accretion rate. 

\section{Black Hole Evolution} \label{BH_evo}
The evolution of the BH's spin is driven by the interplay of hydrodynamic torques from accreting matter and electromagnetic torques from the jet. The infalling material imparts angular momentum to the BH, spinning it up, while the jet extracts angular momentum, exerting a spin-down torque.

\subsection{Standard Thin disk}
For a disk with weak large-scale magnetic flux, the dynamical effects of the magnetic field to the disk strucure become negligible, allowing the use of the standard, {thin}, radiatively efficient Novikov–Thorne (NT) disk model \citep{1973blho.conf..343N}. In this model, gas follows quasi Keplerian orbits, slowly spiraling inward until it reaches the innermost stable circular orbit (ISCO) at radius $R_\mathrm{ms}$, the gas plunges into the event horizon. Then the evolution of the BH's mass and spin follows the Bardeen theory \citep{1970Natur.226...64B}, governed by the specific angular momentum and energy at ISCO
\begin{equation}
\frac{1}{\dot{m}} \frac{\mathrm{d} a}{\mathrm{~d} t} = \frac{s_{\mathrm{NT}}(a)}{M}, \quad \frac{1}{\dot{m}} \frac{\mathrm{d} M}{\mathrm{~d} t} = e_{\mathrm{NT}},
\end{equation}
where the spin-up parameter \(s_{\mathrm{NT}}\) is given by
\begin{equation}
s_{\mathrm{NT}}(a) = l_{\text {in }}(a) - 2a e_{\text {in }}(a)
\end{equation}
while $e_{\mathrm{NT}} = e_{\mathrm{in}}$, $e_{\mathrm{in}}$ and $l_\mathrm{in}$ are the specific energy and angular momentum at ISCO, respectively. The formula of $e_{\text{in}}$ and $l_{\text{in}}$ is given in Appendix~\ref{form_app}.

In the standard thin disk, the angular momentum from matter infalling from the ISCO is always sufficient to spin up the BH, resulting in a positive $s_{\mathrm{NT}}(a)$ that decays to zero only at $a = 1$, where the BH reaches its maximally allowed spin. Given that $L_{\mathrm{BZ}} \propto \Phi_{\mathrm{BH}}^2 \Omega_{\mathrm{H}}^2$ and $\Omega_{\mathrm{H}}$ remains nearly constant once the spin is maximized during GRBs, the jet luminosity and energy are primarily constrained by the magnetic flux $\Phi_{\mathrm{BH}}$.

\subsection{Magnetically Arrested disk}
In the MAD regime, the jet power and electromagnetic torque are coupled with the accretion power through the jet efficiency \(\eta_{\mathrm{EM}}(a)\), which depends on the BH spin \citep{2024ApJ...960...82L}. The spin evolution is described by
\begin{equation}
\frac{1}{\dot{m}} \frac{\mathrm{d} a}{\mathrm{~d} t} = \frac{s_{\mathrm{MAD}}(a)}{M}, \quad \frac{1}{\dot{m}} \frac{\mathrm{d} M}{\mathrm{~d} t} = e_{\mathrm{HD}} - \eta_{\mathrm{EM}}(a),
\end{equation}
where \(s_{\mathrm{MAD}}(a)\) is the spin-up parameter in the MAD state
\begin{equation} \label{s_MAD}
s_{\mathrm{MAD}} = \left(l_{\mathrm{HD}} - 2a e_{\mathrm{HD}}\right) - \eta_{\mathrm{EM}}(a)\left(\frac{c}{k(a) \Omega_{\mathrm{H}} r_\mathrm{g}} - 2a\right).
\end{equation}
Here, \(e_{\mathrm{HD}}\) and \(l_{\mathrm{HD}}\) represent the hydrodynamic energy and angular momentum fluxes onto the BH, respectively, with \(k(a) = \Omega_{\mathrm{F}} / \Omega_{\mathrm{H}}\) being the ratio of the magnetic field line's angular frequency to the BH's horizon angular frequency. The numerical values of $e_{\mathrm{HD}}$ and $l_{\mathrm{HD}}$ are approximated as constants by averaging, with $l_{\mathrm{HD}}=0.86$ and $e_{\mathrm{HD}} = 0.97$, and $\eta_{\mathrm{EM}}(a)$ is given by Equation~\ref{eta_EM}. The function $k(a)$ is given by \citep{2024ApJ...960...82L}
\begin{equation}
k(a) = \left\{
\begin{aligned}
& 0.23, \text{ if } a < 0, \\
& \min(0.1 + 0.5a, 0.35), \text{ if } a \geq 0. 
\end{aligned}
\right.
\end{equation}

In contrast to the standard thin disk, where the BH always spins up over time, the BH can spin down rapidly in a MAD. The MAD supplies significantly less hydrodynamic specific angular momentum than the NT disk, which has a minimum value of \(l_{\text{in}} \sim 1.15\). This is likely due to strong magnetic fields both efficiently transporting angular momentum outward and causing the disk to become sub-Keplerian, thereby starving the BH of hydrodynamic angular momentum \citep{2024ApJ...960...82L}. 

In addition to the magnetic fields extracting angular momentum from the BH gas supply, the magnetic fields threading the BH also extract the angular momentum from the BH and send it out to large distances in the form of jets. For $a \gtrsim 0.1$, the jet torque dominates and brakes the BH efficiently. Both of these effects lead to a low equilibrium BH spin of $a_\mathrm{eq} \approx 0.06$.

\subsection{A Unified Description of the disk}
As the magnetic flux threading the BH approaches the MAD limit, the large-scale magnetic field begins to influence disk dynamics, extracting significant energy and angular momentum from the BH. We introduce the transition function $\Delta$, defined as the ratio of the BH magnetic flux to its MAD value
\begin{equation} 
\Delta = \frac{\Phi_{\mathrm{BH}}}{\Phi_{\mathrm{MAD}}}, 
\end{equation} 
where $\Phi_{\mathrm{MAD}}$ represents the maximum flux retained by accretion \begin{equation} \label{Phi_MAD} 
\Phi_\mathrm{MAD}(a,\dot{m}, M_\mathrm{BH}) = \sqrt{\frac{ 16 \pi }{\kappa f c}} \frac{\sqrt{\eta_{\mathrm{EM}}(a)\dot{m} c^2}}{\omega_\mathrm{H}(a)} r_\mathrm{g}. \end{equation} 
When $\Phi_{\mathrm{BH}} > \Phi_{\mathrm{MAD}}$, the magnetic field rapidly diffuses outward, restoring $\Phi_{\mathrm{BH}}$ to $\Phi_{\mathrm{MAD}}$. Therefore, $\Delta$ ranges from 0 to 1, characterizing the disk's state relative to MAD.

We propose the modified spin and mass evolution equations that account for transitions between the NT disk and MAD, {guaranteed to be accurate in both asymptotic limits}: 
\begin{equation} \label{full_a} 
\frac{1}{\dot{m}} \frac{\mathrm{d} a}{\mathrm{~d} t}=\frac{\Delta s_{\mathrm{MAD}}(a) + s_{\mathrm{NT}}(1-\Delta)}{M}, 
\end{equation} 
\begin{equation} \label{full_M} 
\frac{1}{\dot{m}} \frac{\mathrm{d} M}{\mathrm{~d} t}=e_{\mathrm{NT}}(1-\Delta)+ \Delta \left(e_{\mathrm{HD}}-\eta_{\mathrm{EM}}(a)\right). 
\end{equation} 
These equations recover the standard NT disk behavior for $\Delta \rightarrow 0$ and the MAD limit for $\Delta = 1$. The remaining unknowns, $\dot{m}$ and $\Phi_{\mathrm{BH}}$, along with initial conditions for $M_{\mathrm{BH,0}}$ and $a_0$, fully determine the evolution of the BH and its jet power, depending on the specific astrophysical model.

Additionally, since $L_\mathrm{jet} \propto \Phi_\mathrm{BH}^2$, the jet power can be expressed as
\begin{equation} \label{Power_Delta} 
L_\mathrm{jet} = \Delta^2 \eta_{\mathrm{EM}}(a) \dot{m} c^2. 
\end{equation} 
This expression illustrates the jet power's dependence on the MAD transition.

Finally, the spin equilibrium timescale $t_\mathrm{eq}$ is given by
\begin{equation} \label{eq_time}
\begin{aligned}
    t_{\text{eq}} & \equiv \left|(a_\mathrm{eq}-a_0)/{\frac{da}{dt}\Big|_{a=a_0}} \right|\\
    & \sim \frac{M_0}{\dot{m}} \left|\frac{a_\mathrm{eq}-a_0}{\Delta_0 \, s_{\text{MAD}}(a_0)+(1-\Delta_0)s_{\text{NT}}(a_0)} \right|,
\end{aligned}
\end{equation}
where the factor $\left|\frac{a_\mathrm{eq}-a_0}{\Delta_0 \, s_{\text{MAD}}(a_0)+(1-\Delta_0)s_{\text{NT}}(a_0)}\right| < 0.3$ for all $a_0$ and $\Delta_0$. For typical GRBs, $t_\mathrm{eq} \lesssim 10s$, shorter than the duration of lGRBs but longer than sGRBs. Thus, the spin equilibrium state dominates the observable properties of lGRBs, while the initial spin governs those of sGRBs. We will apply this model to both types of GRBs in subsequent sections separately.

\section{Long Gamma-Ray Bursts} \label{lGRBs}
\subsection{Collapsar Model Setups}
We consider a GRB jet powered by accretion onto a BH following the core collapse of a WR star. The mass accretion rate depends on both the stellar structure and the interaction with relativistic jets. For simplicity, we assume a constant accretion rate throughout the burst, consistent with observations of approximately constant fluence across the sequence of pulses that make up the light curve of a lGRB \citep{2002A&A...393L..29M}. 

Strong accretion retains large-scale vertical magnetic flux on the BH, crucial for the BZ mechanism to power relativistic jets. However, the origin of this magnetic flux remains poorly understood, with several models proposed \citep{2015MNRAS.447..327T,2024MNRAS.532.1522J,2024arXiv240716745G}. Since the flux is expected to be transported and retained by accreted matter, it is reasonable to parametrize it in relation to the total accreted mass, approximated as $M_{\mathrm{BH}}$:
\begin{equation} \label{Phi_BH}
    \Phi_{\mathrm{BH}}(t)=\Phi_{\mathrm{BH,0}}\left(\frac{M_{\mathrm{BH}}(t)}{M_{\mathrm{BH,0}}}\right),
\end{equation}
where $M_{\mathrm{BH,0}}$ is the initial BH mass, and $\Phi_{\mathrm{BH,0}}$ is the initial magnetic flux threading the BH. This parameterization assumes a constant magnetic flux per unit mass during the initial collapse and subsequent accretion, with a uniform magnetic field orientation across all accreted material. Consequently, the total magnetic flux threading the BH is the cumulative sum of the flux contributed by the accreted material and is therefore proportional to the mass of the BH. Further discussion on this assumption is provided in Section~\ref{discussion}.

In the collapsar model, the BH is initially buried within a dense stellar envelope. The jet requires several seconds to drill through the collapsing star, with the GRB triggered when the jet breaks out of the star. We adopt the following expression for the breakout time of a Poynting-flux dominated constant luminosity relativistic jet\footnote{Here we approximate the density profile of the stellar envelope to be $\rho \propto r^{-\alpha}$ with index $\alpha = 2.5$.}
\citep{2014MNRAS.443.1532B,2015MNRAS.450.1077B},
\begin{equation} \label{Tau_break}
\begin{aligned}
\tau_{\text {breakout}} \simeq & \, 7.4 \,\left(\frac{L_{\mathrm{j}}}{ 10^{49} \mathrm{erg} \mathrm{s}^{-1}}\right)^{-1 / 3} \left(\frac{M_*}{15 \mathrm{M}_{\odot}}\right)^{1 / 3}\\
& \times \left(\frac{M_\mathrm{BH}}{3 \mathrm{M}_{\odot}}\right)^{2 / 3} \left(\frac{0.05}{\omega_{\mathrm{H}}(a)}\right)^{2 / 3}[\mathrm{s}]
\end{aligned}
\end{equation}
where $\tau_{\text {breakout}}$ is the time for the jet to traverse the star, minus the light travel time across it. The jet power and BH spin can evolve significantly within the first few seconds, affecting the breakout time. We thus approximate\footnote{Equation~\ref{Tau_break} assumes constant jet luminosity, whereas our model considers time-dependent jet power. Consequently, $\tau_{\text {breakout}}$ depends on the time at which it is estimated, and the evolution time should also be included. Thus, we approximate the breakout time as the minimum of $\left(\tau_{\text {breakout}} + t\right)$ over the evolution. The stellar envelope is assumed to remain largely unchanged until the jet breaks through.} the breakout time $t_\mathrm{b}$ as the global minimum of the function $\tau_{\text {breakout}}(t) + t$.

\begin{figure*}
    \vspace{-0.5cm} 
    \gridline{
        \fig{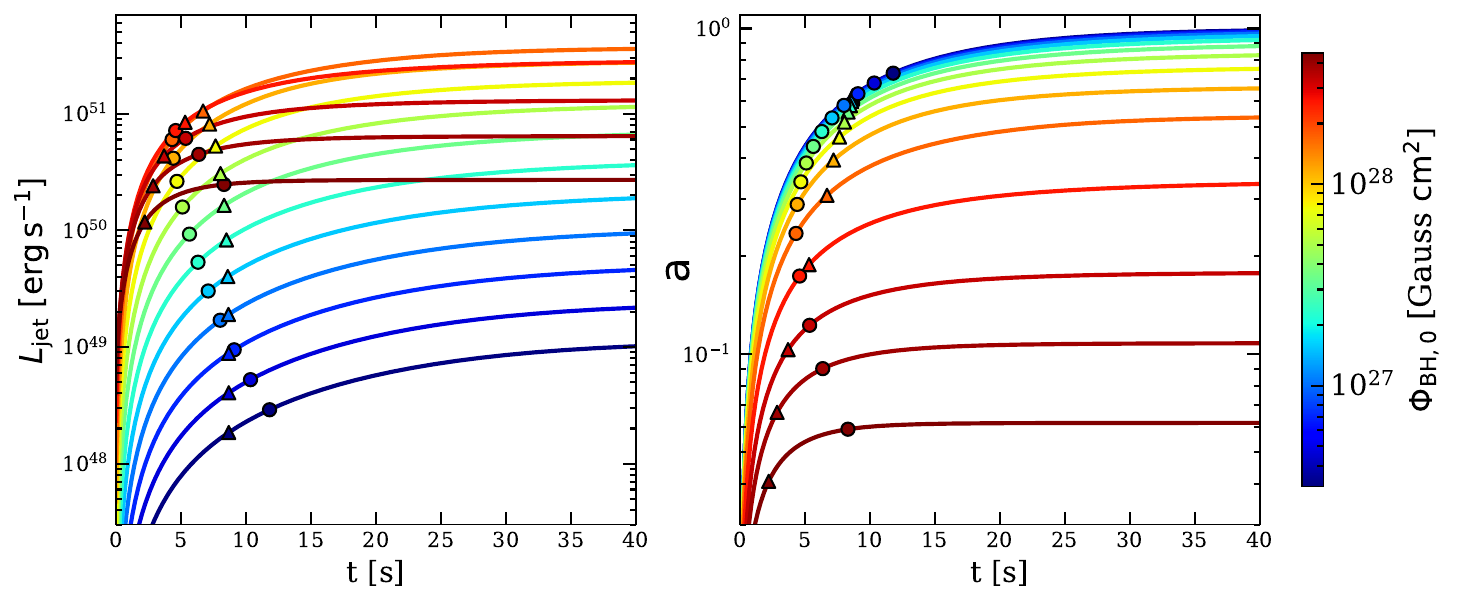}{\textwidth}{}
    }
    \vspace{-0.8cm} 
    \gridline{
        \fig{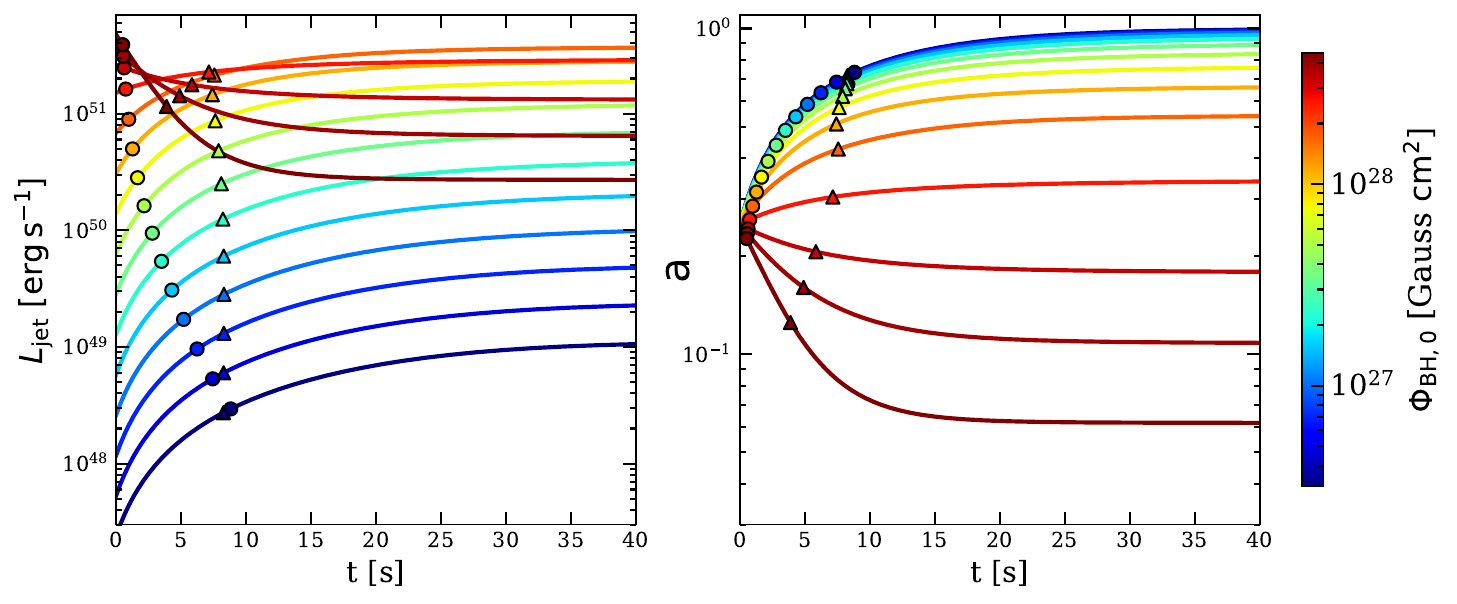}{\textwidth}{}
    }
    \vspace{-0.75cm} 
    \caption{Evolution of jet power and black hole spin for initial magnetic fluxes $\Phi_\mathrm{BH,0} = 10^{26.5-28.65}\,\mathrm{G}\,\mathrm{cm}^2$, initial BH spin $a_0 = 0.01$ (top) and $a_0 = 0.25$ (bottom). The black hole initial mass, accretion rate, and progenitor mass are fixed at $M_{{\mathrm{{BH,0}}}} = 3 M_{{\odot}}$, $\dot{{m}} = 0.1 M_{{\odot}}\,\mathrm{s}^{{-1}}$, and $M_{\ast} = 15 M_\odot$, respectively. Colored dots mark the breakout time $t_\mathrm{b}$ of the jet from the stellar envelope and colored triangles mark the spin equilibrium timescale $t_\mathrm{eq}$ given in Equation~\ref{eq_time}. In both cases, jet power and spin approach equilibrium after $t \sim 10s$. As the magnetic flux approaches the MAD limit, the equilibrium spin decreases, reaching $a_\mathrm{eq} \approx 0.06$ in the fully MAD state. The equilibrium timescale $t_\mathrm{eq}$ is similar to $t_\mathrm{b}$ for $a_0 = 0.01$, but it is significantly longer than $t_\mathrm{b}$ for $a_0 = 0.25$, especially when the magnetic flux is high. The relatively constant observed jet luminosity suggests a small initial black hole spin in collapsar models.}
    \label{fig:L_time}
\end{figure*}

Figure~\ref{fig:L_time} shows the evolution of jet power and BH spin for initial magnetic fluxes $\Phi_\mathrm{BH,0} = 10^{26.5-28.65}\,\mathrm{G}\,\mathrm{cm}^2$ with initial BH spin $a_0 = 0.01$ (top) and $a_0 = 0.25$ (bottom). For a typical WR star, $M_\ast \sim 15 M_\odot$ and $R_\ast \sim 0.5 R_\odot$, leading to $\dot{{m}} \sim 0.1M_{{\odot}}\mathrm{s}^{{-1}}$ via free-fall collapse. In both cases, jet power and spin approach equilibrium after $t \sim 10s$. As the magnetic flux approaches the MAD limit, the equilibrium spin decreases, reaching $a_\mathrm{eq} \approx 0.06$ in the fully MAD state. The colored circles mark the jet breakout time $t_\mathrm{b}$, after which the GRB prompt emission becomes observable, and the colored triangles mark the spin equilibrium timescale $t_\mathrm{eq}$ given in Equation~\ref{eq_time}.

The collimation-corrected jet power for typical lGRBs is $L_\mathrm{jet} \sim 10^{50}\, \mathrm{erg}\,\mathrm{s}^{-1}$ \citep{2010MNRAS.406.1944W,2018ApJ...859..160W}, corresponding to a breakout time of $t_\mathrm{b} \sim 8\, \mathrm{s}$ for $a_0 = 0.01$ in Figure~\ref{fig:L_time}. This breakout time aligns with the observed plateau in the duration distribution \citep{2012ApJ...749..110B}, and the roughly constant jet power beyond $t_\mathrm{b}$ is consistent with typical lGRB observations \citep{2002A&A...393L..29M}. Additionally, Equation~\ref{eq_time} yields $t_\mathrm{eq} \lesssim 9 \,\mathrm{s}$ for $M_0 = 3 M_\odot$, $\dot{{m}} = 0.1 M_{{\odot}}s^{{-1}}$ and $a_0 = 0.01$. Thus, the system typically approaches the spin equilibrium state soon after the jet emerges.

In contrast, the bottom panel shows significantly shorter breakout times for $a_0 = 0.25$, as the higher initial spin and jet power allow faster penetration of the stellar envelope. However, this rapid breakout and subsequent power decay due to spin-down are inconsistent with observed lGRB behavior, suggesting that models assuming moderate or high initial BH spin with strong magnetic flux are less plausible for collapsars. Similar conclusions were drawn by \cite{2024ApJ...961..212J} with different models for the jet breakout time.
Interestingly, the harder spectra observed in the first 1–2 seconds of some long GRBs resemble those of short GRBs \citep{2009A&A...496..585G}. This similarity could arise if the spin equilibrium time slightly exceeds the breakout time, allowing the BH spin to evolve during the initial emission phase, similar to an sGRB. 

\subsection{Spin Equilibrium and Jet Energetics}
Since we can only observe the GRB prompt emission after the jet emerges from the stellar envelope, and by that time the BH spin does not evolve substantially, then the spin equilibrium state plays a crucial role in determining lGRB observables. As implied in Figure~\ref{fig:L_time}, the equilibrium spin $a_\mathrm{eq}$ depends on the transition function at spin equilibrium, $\Delta_\mathrm{eq}$, and the dependence can be derived from Equation~\ref{full_a}.

The transition function $\Delta$ is the ratio between $\Phi_\mathrm{BH}(M_\mathrm{BH})$ and $\Phi_\mathrm{MAD}(a,\dot{m}, M_\mathrm{BH})$. With a constant mass accretion rate, $\Delta_\mathrm{eq}$ is determined solely by the equilibrium spin $a_\mathrm{eq}$ as the mass is canceled out, leading to a direct one-to-one mapping between $\Delta_\mathrm{eq}$ and $a_\mathrm{eq}$. The spin equilibrium is achieved once the numerator on the right-hand side of Equation~\ref{full_a} vanishes. Then the jet power at spin equilibrium is given by $L_\mathrm{jet,eq} = \Delta_\mathrm{eq}^2 \eta_\mathrm{EM}(\Delta_\mathrm{eq}) \dot{m} c^2$. $M_\mathrm{BH,0}$ and $\dot{m}$ only determine how fast the equilibrium is reached. 

\begin{figure}
    \centering
    \includegraphics[width=\columnwidth]{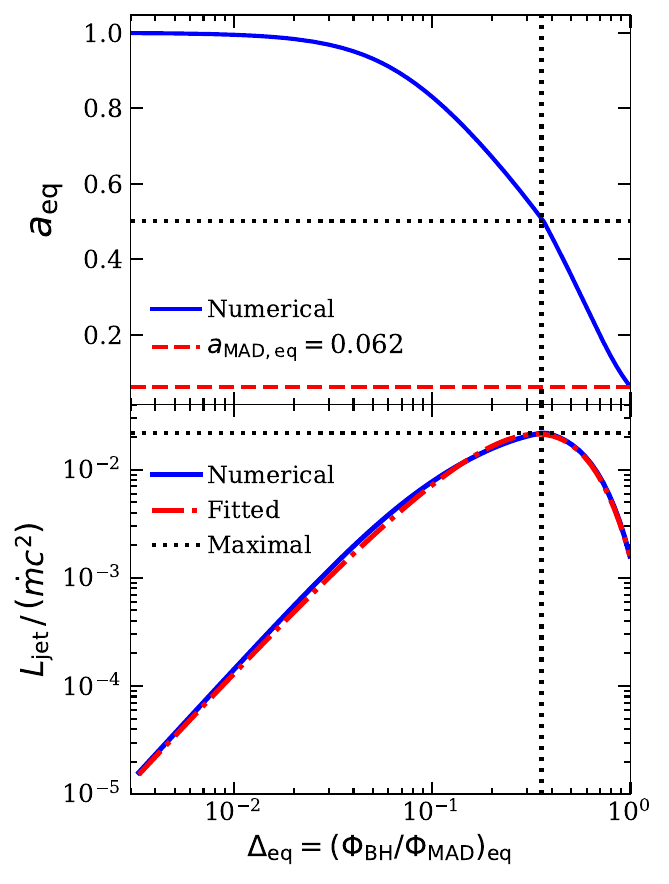}
    \caption{Top panel: Equilibrium spin $a_\mathrm{eq}$ as a function of the transition parameter at spin equilibrium, $\Delta_\mathrm{eq} = \left(\Phi_\mathrm{BH}/\Phi_{\mathrm{MAD}}\right)_\mathrm{eq}$. The dashed red line marks the equilibrium spin at the MAD limit. Bottom panel: Normalized spin-equilibrium jet power, $L_\mathrm{jet,eq}/(\dot{m} c^2)$, as a function of $\Delta_\mathrm{eq}$. The solid blue lines represent the full numerical calculations, while the red dash-dotted line shows the fitted results using Equation~\ref{fitting_form}. The black dotted lines indicate the maximum jet power at spin equilibrium, which is $2.2\%$ of the accretion power $\dot{m}c^2$, corresponding to $\Delta_\mathrm{eq,max} = 0.365$ and $a_\mathrm{eq,max} = 0.5$. The maximum spin-equilibrium jet power is approximately 15 times greater than that at the MAD limit, which is 0.16\% of the accretion power. }
    \label{fig:L_Delta}
\end{figure}

Figure~\ref{fig:L_Delta} illustrates $a_\mathrm{eq}$ and the normalized jet power $L_\mathrm{jet,eq}/(\dot{m} c^2)$ as functions of $\Delta_\mathrm{eq}$. When $\Delta_\mathrm{eq} \ll 1$, the system approaches the standard thin disk limit with $a_\mathrm{eq} \sim 1$. In contrast, $\Delta_\mathrm{eq} = 1$ corresponds to the MAD state, where $a_\mathrm{eq} \sim 0.06$. At the standard thin disk limit, $L_\mathrm{jet,eq}/(\dot{m} c^2) \sim 1.46 \Delta_\mathrm{eq}^2$, which matches a power-law behavior of index 2. This motivates the following fitting form: \begin{equation} \label{fitting_form} \frac{L_\mathrm{jet,eq}}{\dot{m} c^2} = 1.458 \Delta_\mathrm{eq}^{2 + C_0 \Delta_\mathrm{eq}} \exp \left(-\left( \frac{\Delta_\mathrm{eq}}{C_1} \right)^{C_2} \right), \end{equation} with $C_0 = 2.3$, $C_1 = 0.3$, and $C_2 = 1.6$, shown as the red dash-dotted line in Figure~\ref{fig:L_Delta}.

For a constant $\dot{m}$, the maximum jet power at spin equilibrium is $2.2\%$ of $\dot{m} c^2$, occurring at $\Delta_\mathrm{eq,max} = 0.365$ and $a_\mathrm{eq,max} = 0.5$. In contrast, at the MAD limit, jet power is only $0.16\%$ of $\dot{m} c^2$, approximately 14 times lower. This maximum power occurs midway between the standard thin disk and MAD states, where moderate spin and magnetic flux combine to produce the most powerful jets for a given accretion rate. 

\begin{figure}
    \centering
    \includegraphics[width=\columnwidth]{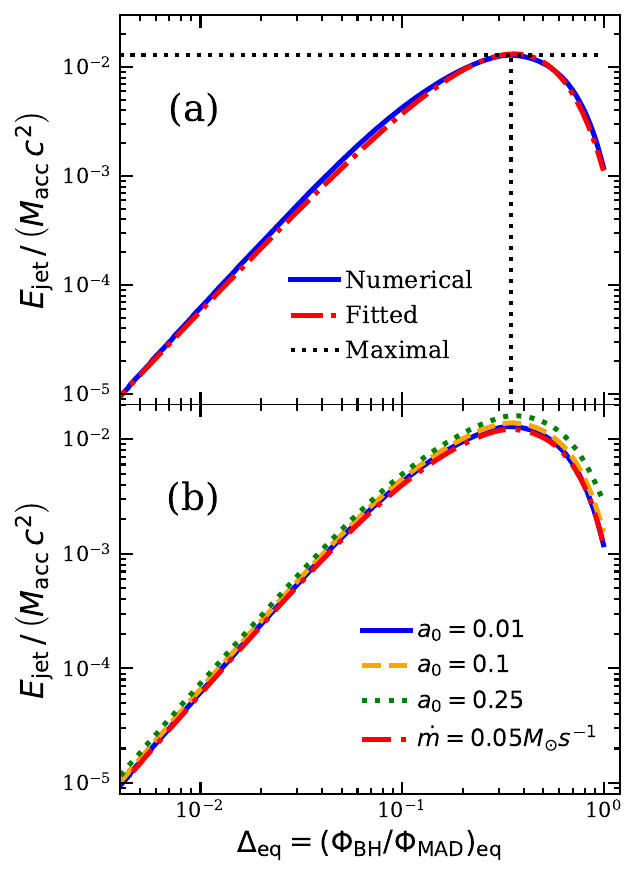}
    \caption{(a) The observed cumulative energy of the jet normalized by the total accretion energy, $E_\mathrm{jet}/(M_\mathrm{acc} c^2)$, as a function of the transition function at spin equilibrium $\Delta_\mathrm{eq} = \left(\Phi_\mathrm{BH}/\Phi_{\mathrm{MAD}}\right)_\mathrm{eq}$. The solid blue line shows the full numerical calculations for $\dot{m} = 0.1 M_\odot\, \mathrm{s}^{-1}, t_\mathrm{e} = 40\,\mathrm{s}, a_0 = 0.01, M_\mathrm{BH,0} = 3 M_\odot, M_\ast = 15 M_\odot$, while the red dash-dotted line represents the fit from Equation~\ref{fitting_form_Ejet}. The black dotted lines mark the maximum observed cumulative jet energy, $E_\mathrm{jet,max}$, which is 1.44\% of the accretion energy, about 12 times higher than that for the MAD state. (b) The blue solid curve is the same as (a), while the orange dashed and green dotted lines show results for different initial spins, $a_0 = 0.1$ and $a_0 = 0.25$, respectively. The red dash-dotted line shows results for $\dot{m} = 0.05 M_\odot\,\mathrm{s}^{-1}$ and $t_\mathrm{e} = 80\,\mathrm{s}$, keeping the total accreted mass $M_\mathrm{acc}$ constant. For $a_0 \lesssim 0.25$, the normalized jet energy is largely insensitive to system parameters. }
    \label{fig:E_normalized}
\end{figure}

The observed cumulative jet energy, $E_\mathrm{jet}$, is another key parameter and is given by \begin{equation} E_\mathrm{jet} = \int_{t_\mathrm{b}}^{t_\mathrm{e}} L_\mathrm{jet} \, dt \approx L_\mathrm{jet,eq} \left( t_\mathrm{e} - t_\mathrm{b} \right) \approx L_\mathrm{jet,eq} t_\mathrm{e}, \end{equation} since $t_\mathrm{b}$ is small compared to the total engine activity time $t_\mathrm{e}$ in most lGRBs. Defining the total accreted mass as $M_\mathrm{acc} = \dot{m} t_\mathrm{e}$, the normalized cumulative jet energy $E_\mathrm{jet}/(M_\mathrm{acc} c^2)$ closely follows the normalized jet power $L_\mathrm{jet,eq}/(\dot{m} c^2)$, as confirmed by Figure~\ref{fig:E_normalized}.\footnote{Reducing the total engine activity time from $t_\mathrm{e}=40\,\mathrm{s}$ to lower values (e.g., $t_\mathrm{e} = 20\,\mathrm{s}$) results in only minor changes, with variations of the curve within a factor of two.}

Therefore, we propose a similar fitting form for the normalized cumulative jet energy: \begin{equation} \label{fitting_form_Ejet} \frac{E_\mathrm{jet}}{M_\mathrm{acc} c^2} = C\Delta_\mathrm{eq}^{(2 + C_0\Delta_\mathrm{eq})}\exp\left(-\left(\frac{\Delta_\mathrm{eq}}{C_1}\right)^{C_2}\right), \end{equation} with the fitting parameters $C = 0.62$, $C_0 = 1.47$, $C_1 = 0.31$, and $C_2 = 1.58$. The primary changes are in the normalization and power-law index, due to the influence of the non-negligible jet breakout time, which increases significantly when the magnetic flux is small.

The horizontal black dotted line in Figure~\ref{fig:E_normalized} marks the maximum observed cumulative jet energy, $E_\mathrm{jet,max}$, which is 1.44\% of the accretion energy, 12 times greater than the 0.12\% for the MAD state. The efficiency is lower than the one associated with the maximum jet power, as part of the energy is expended for the jet to break through the stellar envelope. The corresponding flux ratio at spin equilibrium is $\Delta_\mathrm{eq,max} \sim 0.36$, shown by the vertical black dotted line. While some non-equilibrium jet power is included, the results remain largely insensitive to system parameters for initial spins $a_0 < 0.1$. For $a_0 \gtrsim 0.25$, deviations occur due to smaller $t_\mathrm{b}$, particularly at high magnetic flux. As discussed earlier and supported by other studies \citep{2019ApJ...881L...1F,2023ApJ...952L..32G}, collapsar BHs are likely born with low spins, especially when the initial flux is high.  Therefore, this curve of normalized jet energy applies universally across a broad range of lGRBs.

\subsection{Comparison with Observations}
We focus on the collimation-corrected jet energy, $E_\mathrm{jet}$, which is less affected by system fluctuations and observational uncertainties compared to jet power. Although $M_\mathrm{acc}$ is not directly observable, we can still compare $E_\mathrm{jet}$ with observations to reveal meaningful insights.

\subsubsection{Standard lGRBs}
\begin{figure*}
    \centering
    \includegraphics[width=0.95\textwidth]{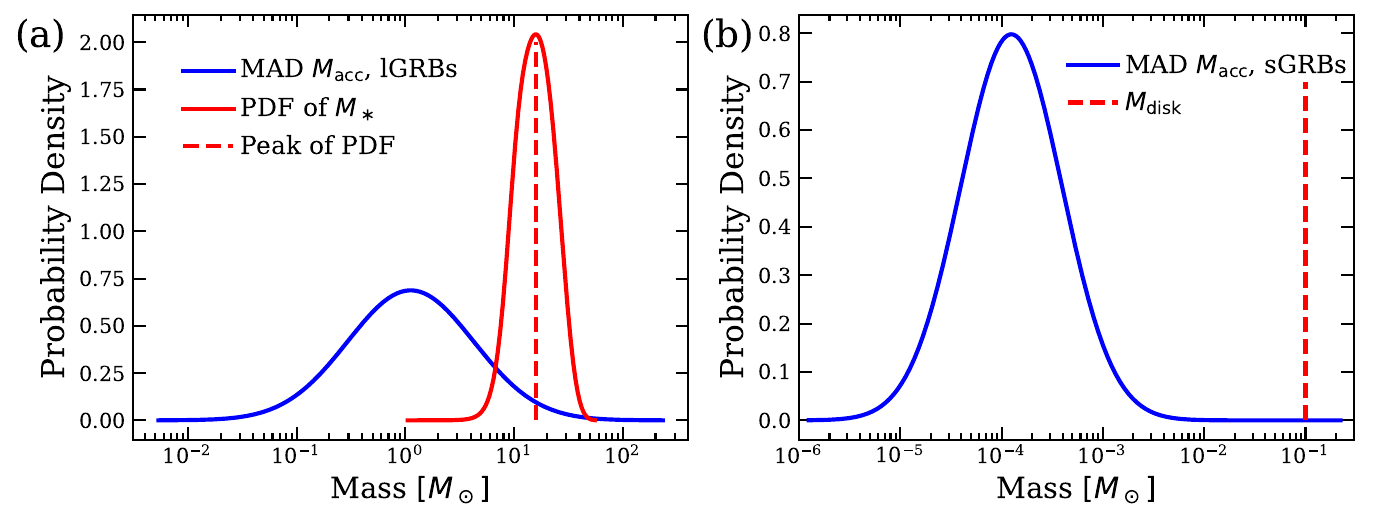}
    \caption{(a) Distribution of accreted mass, $M_\mathrm{acc}$, assuming all lGRBs are produced in the MAD limit (blue solid line). Red solid line represents the mass distribution of Galactic WC and WO stars \citep{2019A&A...621A..92S} with the red dashed line indicating the peak. If the MAD assumption holds, the $M_\mathrm{acc}$ distribution should peak around 10\% of the $M_\ast$ peak. However, for subluminous lGRBs, $M_\mathrm{acc}$ may be as low as $\sim 0.1 M_\odot$, potentially too small relative to the progenitor mass. Additionally, the broad distribution of accreted masses does not align with the narrow range of progenitor masses. (b) Accreted mass, $M_\mathrm{acc}$, distribution for sGRBs under the MAD limit (blue solid line). The red dashed line indicates the typical mass of the remnant disk ($\sim 0.1 M_\odot$). The $M_\mathrm{acc}$ distribution peaks at around 0.1\% of the disk mass, suggesting that the accreted mass is too small for typical remnant disks. The tensions in (a) and (b) can be alleviated if lGRBs and standard sGRBs are produced by systems far from the MAD state.}
    \label{fig:M_acc}
\end{figure*}

For lGRBs of typical energy, assuming they are produced by systems in the MAD limit  leads to potential inconsistencies between the mass accretion predicted by the MAD model and that required by observations. Figure~\ref{fig:M_acc}(a) shows the predicted distribution of $M_\mathrm{acc}$ if all lGRBs were produced by MAD systems, where the peak accreted mass would be around 10\% of the progenitor mass. However, for subluminous lGRBs, $M_\mathrm{acc}$ would need to be as low as $\sim 0.1 M_\odot$, which may be too small relative to the progenitor star's mass. Additionally, it is difficult to reconcile how a narrow distribution of Galactic WC and WO star masses \citep{2019A&A...621A..92S} could produce a broad distribution of accreted masses without additional influencing factors.

This discrepancy can be alleviated if subluminous lGRBs have $\Delta_\mathrm{eq} \ll 1$, which would imply a larger $M_\mathrm{acc}$ for a given $E_\mathrm{jet}$, as shown in Figure~\ref{fig:lognorm}. For standard GRBs, if they are not produced by MAD, the available magnetic flux in the system provides an additional factor that may explain the observed diversity in GRB jet energies. Consequently, it is likely that most lGRBs are generated in systems far from MAD. In this scenario, with $a_\mathrm{eq} \approx 1$, the jet luminosity would primarily depend on the magnetic flux. With certain additional assumptions, the distributions of $\Delta_\mathrm{eq}$ and $\Phi_\mathrm{BH,eq}$ for a fixed $M_\mathrm{acc}$ are provided in Appendix~\ref{distributions}. {The inferred magnetic flux distribution peaks at around $\sim 10^{27} \text{ G cm}^2$ and remains largely unaffected by variations in the assumed $M_\mathrm{acc}$.}

\subsubsection{The BOAT}
\begin{figure}
    \centering
    \includegraphics[width=\columnwidth]{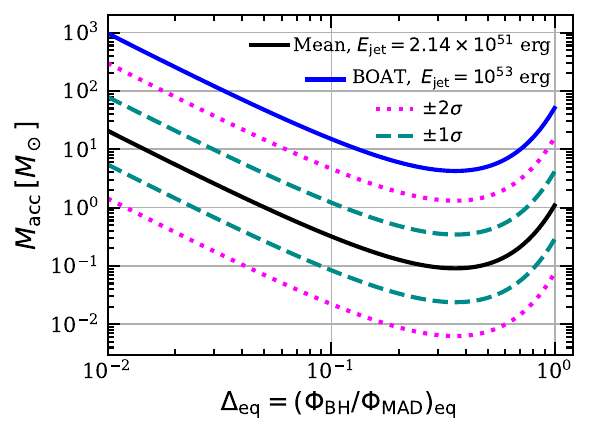}
    \caption{Mean collimation-corrected jet energy inferred from observations \citep{2018ApJ...859..160W} (black solid line). Dark cyan dashed and magenta dotted lines show the one and two $\sigma$ confidence intervals, respectively. The blue solid line indicates the energy of the BOAT. If the BOAT were produced by a system in the MAD state, the required accreted mass would reach $M_\mathrm{acc} \sim 50 M_\odot$. In contrast, with the optimal flux ratio ($\Delta_\mathrm{eq} = \Delta_\mathrm{eq,max}$), $M_\mathrm{acc}$ is reduced to $\sim 4 M_\odot$.}
    \label{fig:lognorm}
\end{figure}

GRB 221009A, referred to as the Brightest Of All Time (BOAT; \citealt{2023ApJ...946L..31B}), has an isotropic-equivalent energy of $E_{\gamma,\mathrm{iso}} = (1.01 \pm 0.007) \times 10^{55}\, \mathrm{erg}$ \citep{2023ApJ...952L..42L}. While there is ongoing debate about the jet structure and opening angle, the collimation-corrected energy is estimated at $E_\mathrm{jet} \sim 10^{53} \, \mathrm{erg}$ \citep{2023ApJ...948L..12K, 2023SciA....9I1405O}. For comparison, the distribution of $\log_{10}(E_\mathrm{jet})$ can be approximated as Gaussian with a mean of $51.33$ and a standard deviation of $0.58$ \citep{2018ApJ...859..160W}.

Figure~\ref{fig:lognorm} shows the mean and standard deviations of $E_\mathrm{jet}$ in the $M_\mathrm{acc}$ and $\Delta_\mathrm{eq}$ plane, with the BOAT energy also indicated. Each line corresponds to a fixed $E_\mathrm{jet}$ and is the reciprocal of the lines in Figure~\ref{fig:E_normalized}. In the case of the optimal flux ratio ($\Delta_\mathrm{eq} = \Delta_\mathrm{eq,max}$), the required $M_\mathrm{acc}$ for the BOAT is reduced to $\sim 4 M_\odot$, whereas the MAD limit requires $M_\mathrm{acc} \sim 50 M_\odot$, which is inconsistent with WR star masses.

\section{Short Gamma-Ray Bursts} \label{sGRBs}
We consider a sGRB jet powered by the accretion of a remnant disk onto a newly formed BH after a NS-NS merger. The initial spin and mass of BHs in this model are relatively well-constrained, with $a_0 \sim 0.7$ and $M_\mathrm{BH,0} \sim 2.5 M_\odot$, based on theoretical expectations from NS mergers with $M_{\mathrm{NS}} \sim 1.4 M_{\odot}$ \citep{Kastaun2013, Bernuzzi2014,2020MNRAS.497.1488B}. A remnant disk forms around the compact object with a typical mass of approximately $0.1 M_\odot$ \citep{2020MNRAS.497.1488B}. Given a typical duration of $T_{90}\sim 0.2\,\mathrm{s}$ for sGRBs \citep{2018Ap&SS.363..223Z}\footnote{Here and throughout, all instances of $T_{90}$ refer to the intrinsic burst duration in the central engine frame.}, the mass accretion rate, assuming a constant $\dot{m}$, is about $0.5 M_\odot s^{-1}$ if most of the disk is accreted during the burst.

Since the accreted mass is much less than that of the BH, its spin remains nearly constant throughout the burst. While the magnetic flux available for jet launching in sGRBs is uncertain, it can be inferred from the observed jet luminosity or energy. Given the nearly constant jet luminosity, the jet energy can be estimated as $E_{\mathrm{jet}}=L_{\mathrm{jet}}T_\mathrm{90}$.

Determining collimation-corrected jet energy for sGRBs is challenging due to their faint afterglows. We assume a Gaussian distribution for $\log_{10}(E_\mathrm{jet})$ with a mean of 50 and a standard deviation of 0.5. Figure~\ref{fig:M_acc}(b) illustrates the predicted $M_\mathrm{acc}$ distribution if all sGRBs are produced in the MAD regime, indicating that the accreted mass is extremely small compared to the disk mass, with the $M_\mathrm{acc}$ distribution peaking around 0.1\% of the disk mass. This is because the jet efficiency is too high in short GRBs, given the high initial spin, suggesting that they are more likely produced by systems outside the MAD regime. The conclusion holds across reasonable choices of $\log_{10}(E_\mathrm{jet})$ distributions, with further details on $\Phi_\mathrm{BH}$ distributions for various assumed $E_\mathrm{jet}$ distributions provided in Appendix~\ref{distributions}. {Similarly to those found for lGRBs, the magnetic flux distributions for sGRBs peak around $\sim 10^{27} \text{ G cm}^2$ with weak dependence on the assumed $E_\mathrm{jet}$ distributions.}

\section{Discussion} \label{discussion}
Given the similar observed isotropic luminosities \citep{2015MNRAS.451..126S} and high BH spin values in both long and short GRBs implied from our model, the distributions of magnetic flux threading the BHs are expected to be similar\footnote{Even though the slightly wider jet opening angles in sGRBs require somewhat smaller collimation correction in comparison to that of lGRBs. \citep{2019MNRAS.483..840B,2023ApJ...959...13R}.}. This suggests similar values and, possibly a common origin, for the magnetic flux that powers relativistic jets in all GRBs. Despite differences in progenitor systems and pre-burst dynamics, a short-lived, rapidly spinning, strongly magnetized proto-neutron star (PNS) may be the essential shared characteristic. Following the PNS collapse, the resulting BH inherits the magnetic field, with strong accretion anchoring this field to the BH \citep{2024arXiv240716745G}. Consequently, the BH forms with strong magnetic fields already threading through it, and the magnetic flux remains nearly constant held in place by the ram pressure of the surounding disk\footnote{In our model, $\Phi_\mathrm{BH}$ also stays nearly constant, as $M_\mathrm{BH}$ does not change significantly during the burst.} throughout the burst.

In lGRBs, the BH may acquire magnetic flux directly through accretion from the progenitor’s envelope \citep{2015MNRAS.447..327T}. However, \cite{2024arXiv240716745G} argue that, even with a strong progenitor magnetic field, its predominantly toroidal, small-scale, and random nature---as predicted by the Tayler-Spruit dynamo \citep{2002A&A...381..923S}---significantly limits its contribution to the net magnetic flux onto the BH after collapse. Similarly, in sGRBs, the remnant disk is expected to have a primarily toroidal field due to tidal disruption and flux freezing \citep{2014PhRvD..90d1502K}. Although this toroidal field may initiate magnetorotational instability, producing weak poloidal fields, it generates jets potentially  too weak to account for GRB luminosities \citep{2012MNRAS.423.3083M}. Therefore, a short-lived, rapidly spinning PNS is likely to serve as the primary source of magnetic flux for both long and short GRBs.

Our model predicts that the maximum observed jet energy is about 1.5\% of the accretion energy, expressed as
\begin{equation}
    E_\mathrm{jet,max} = 1.3\times10^{53}  \left( \frac{M_\mathrm{acc}}{5 M_\odot}\right) \, \mathrm{erg}.
\end{equation}
For comparison, the maximum jet energy from magnetars, even with accretion-induced spin-up, is limited to $3 \times 10^{52} \, \mathrm{erg}$ \citep{2014MNRAS.443...67M}. GRB jets can also be powered by neutrino annihilation, with the jet power\footnote{The expression assumes $a=0.95$ with $\dot{m}_{\mathrm{ign}}<\dot{m}<\dot{m}_{\text {trap }}$, which covers typical GRB accretion rates. However, $L_\mathrm{jet, \nu \bar{\nu}} \sim 0$ once $\dot{m} < \dot{m}_{\mathrm{ign}}$, with $\dot{m}_{\mathrm{ign}} = 0.021 M_{\odot} \,\mathrm{s}^{-1}$ for $a=0.95$ and $\alpha = 0.1$. See Equation 22 in \cite{2011MNRAS.410.2302Z} for details.} given by \citep{2011MNRAS.410.2302Z}
\begin{equation}
L_\mathrm{jet, \nu \bar{\nu}} \sim  7.3 \times 10^{49} \left(\frac{M_{\mathrm{BH}}}{3 M_{\odot}}\right)^{-3 / 2} \left(\frac{\dot{m}}{0.1 M_\odot/\mathrm{s}}\right)^{9 / 4} \mathrm{erg \, \mathrm{s}^{-1}}.
\end{equation}
The neutrino annihilation process can coexist with the BZ mechanism, jointly driving relativistic GRB jets. However, the jet power from neutrino annihilation decreases rapidly as the accretion rate drops. For long-duration GRBs, sustaining the high accretion rates required for neutrino annihilation becomes unfeasible, especially for the energetic ones. For models relying solely on neutrino annihilation, and assuming the final BH mass is twice its initial value, the maximum jet energy is $E_\mathrm{jet,max} \sim 5 \times 10^{51} \, \mathrm{erg}$ for $t_\mathrm{e} \sim 10 \, \mathrm{s}$, with the limit decreasing steeply for longer bursts \citep{2014MNRAS.445L...1L}. With $E_\mathrm{jet} \sim 10^{53} \, \mathrm{erg}$ and $T_\mathrm{90} \sim 250 \, \mathrm{s}$, the BOAT exceeds the energy limits of magnetar\footnote{While some equations of state may allow for more massive, faster-spinning magnetars, even this is unlikely to provide sufficient energy for the BOAT, as not all rotational energy can contribute to the GRB. Specifically, the energy per baryon must be high enough for the outflow to produce a relativistic jet; otherwise, this energy will be wasted.} and pure neutrino annihilation models but can be explained by the BZ mechanism, requiring a reasonable accreted mass of $M_\mathrm{acc} \sim 4 M_\odot$ with optimal magnetic flux.

Note that for BZ powered jets, as studied here, the maximum jet energy for a given accreted mass depends on the chosen transition function, which models the torques exerted on the BH at the transition from the thin disk to MAD. While the flux ratio between the physical and MAD magnetic flux is a natural choice as it approaches the correct limits, alternative options remain. Future 3D GRMHD simulations of sub-MAD state systems, together with calculations of the torque on the BH \citep{2024ApJ...960...82L}, could help validate this choice.

Another caveat of this work is the assumption of a constant mass accretion rate during the GRB prompt emission phase. Note that GRB prompt emission exhibits variability as pulses in the light curve on timescales of seconds to subseconds \citep{2023A&A...671A.112C}, possibly caused by jet energy dissipation processes (e.g., magnetic reconnection) or small-scale fluctuations in the accretion rate. Even in the latter case, we confirm that these fluctuations have a negligible impact on the spin and mass evolution. Significant deviations in jet luminosity due to varying accretion rates arise only when the system approaches the MAD state. Since our analysis focuses on jet energy, these small-scale luminosity variations effectively cancel out. Moreover, when the disk is far from the MAD state, as is likely for most GRBs, the observable jet luminosity\footnote{For lGRBs, the equilibrium jet luminosity dominates the observable jet luminosity, while for sGRBs, the initial luminosity dominates. Both are independent of the accretion rate in this regime.} becomes independent of the accretion rate. Therefore, as long as the accretion rate remains high enough to sustain the magnetic flux and allow equilibrium to be approximately achieved before jet breakout in lGRBs, this assumption does not affect our results.

Our work predicts jet energy, whereas observed GRB data typically provide isotropic equivalent energy. To compare with our predictions, collimation corrections are required, but these depend on the specific jet structure {and identification or limits on the jet break time}, which complicates direct energy estimates \citep{2023SciA....9I1405O}. A detailed and comprehensive analysis of observational data is necessary to accurately determine the jet energy distribution. Future telescopes like LSST \citep{2019ApJ...873..111I} could improve this effort by detecting faint jet breaks with greater sensitivity.
\\
\vspace{1em}

\section{Conclusion} \label{conclude}
In this \textit{letter}, we present a semi-analytical model to explore the evolution of BH central engines and GRB jets over a wide range of magnetic fluxes, allowing transitions from standard thin disk to MAD. We identify a universal curve for normalized jet power $L_\mathrm{jet,eq}/(\dot{m} c^2)$ as a function of the flux ratio at spin equilibrium, $\Delta_\mathrm{eq} = (\Phi_\mathrm{BH}/\Phi_\mathrm{MAD})_\mathrm{eq}$. The maximum jet power reaches $\sim 2.2\%$ of the accretion power $\dot{m} c^2$, occurring at $\Delta_\mathrm{eq,max} = 0.365$ with $a_\mathrm{eq,max} = 0.5$.

For lGRBs, this curve suggests a maximum jet energy of around 1.5\% of the accretion energy---about 12 times higher than that in the MAD state---reducing the required accreted mass for the BOAT from $M_\mathrm{acc} \sim 50 M_\odot$ to $M_\mathrm{acc} \sim 4 M_\odot$. Additionally, the narrow mass distribution of Galactic WC and WO stars suggests that standard lGRBs are unlikely to be produced by the MAD state, given their wide jet energy distribution. Likewise, the energetics and extremely low accreted mass needed for sGRBs indicate they should be produced outside the MAD limit. The magnetic flux distributions for both long and short GRBs peak around $10^{27} \text{ G cm}^2$, suggesting a potentially shared origin of magnetic flux in the central engine in both types of bursts.

\section*{Acknowledgments}
This work was supported by the NSF AST-2308090 grants, a grant (no. 2020747) from the United States-Israel Binational Science Foundation (BSF), Jerusalem, Israel (PB) and by a grant (no. 1649/23) from the Israel Science Foundation (PB). This research has made use of NASA's Astrophysics Data System Bibliographic Services.

\appendix
\twocolumngrid
\section{The formulae for \lowercase{$e_{\text{in}}$} and \lowercase{$l_{\text{in}}$}} \label{form_app}
The formula of specific energy $e_{\text{in}}$ and angular momentum $l_{\text{in}}$ at ISCO are given by \citep{1970Natur.226...64B,1996MNRAS.283..854M} 
\begin{equation}
e_{\text{in}} = \left(1 - \frac{2}{3 R_{ms}}\right)^{1 / 2}, l_{\text {in }} = \frac{2}{3^{3 / 2}}\left(1 + 2\left(3 R_{ms} - 2\right)^{1 / 2}\right),
\end{equation}
while the ISCO radius \(R_{ms}\) can be calculated as
\begin{equation}
R_{ms} = 3 + Z_2 - \left[\left(3 - Z_1\right)\left(3 + Z_1 + 2Z_2\right)\right]^{1 / 2},
\end{equation}
with \(Z_1\) and \(Z_2\) defined by
\begin{equation}
\begin{aligned}
    & Z_1 = 1 + \left(1 - a^2\right)^{1/3}\left[(1 + a)^{1/3} + (1 - a)^{1/3}\right], \\
    & Z_2 = \left(3a^2 + Z_1^2\right)^{1/2}.
\end{aligned}
\end{equation}

\section{magnetic flux and flux ratio distributions} \label{distributions}
\begin{figure*}
    \centering
    \includegraphics[width=\textwidth]{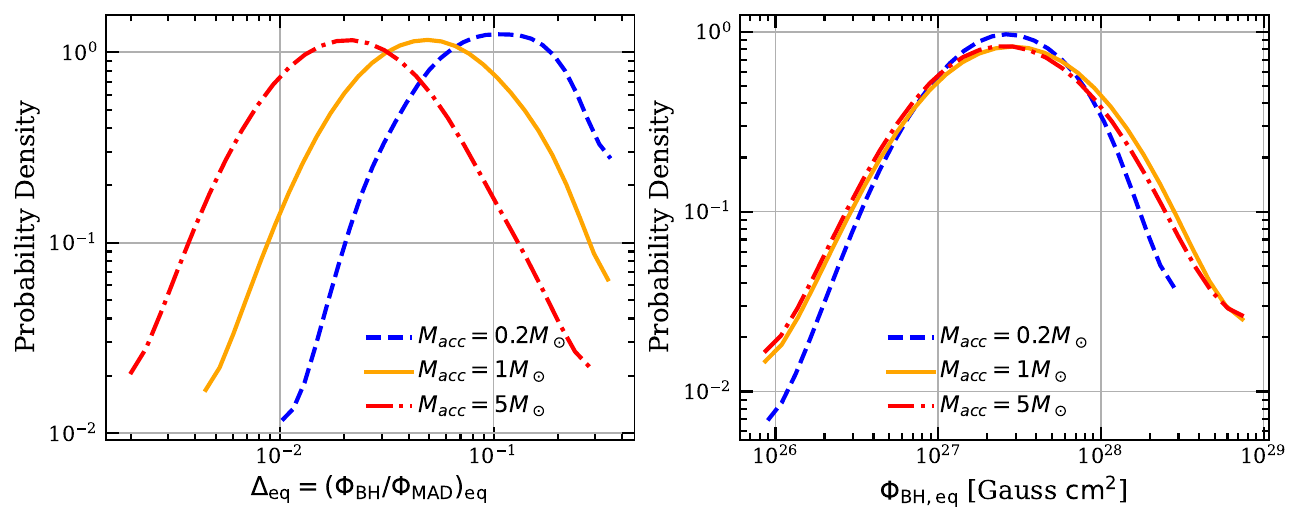}
    \caption{Probability distributions for flux ratio, $\Delta_\mathrm{eq} = (\Phi_\mathrm{BH}/\Phi_\mathrm{MAD})_\mathrm{eq}$ (left), and magnetic flux, $\Phi_\mathrm{BH,eq}$ (right), at spin equilibrium for lGRBs. The blue dashed, orange solid, and red dash-dotted lines correspond to $M_\mathrm{acc} = 0.2, 1, 5 M_\odot$, respectively. As $M_\mathrm{acc}$ increases, systems move further from the MAD state. Even at $M_\mathrm{acc} = 0.2 M_\odot$, the peak of $\Delta_\mathrm{eq}$ remains below $\Delta_\mathrm{eq,max}$, consistent with the assumption that $\Delta_\mathrm{eq} < \Delta_\mathrm{eq,max}$ for all systems. The distributions of $\Phi_\mathrm{BH,eq}$ remain similar across $M_\mathrm{acc}$ values since $a_\mathrm{eq} \sim 1$ in systems far from MAD, and the magnetic flux just follows the observed luminosity distribution.}
    \label{fig:phi_dis}
\end{figure*}

To determine the distribution of magnetic flux $\Phi_\mathrm{BH,eq}$ and flux ratio $\Delta_\mathrm{eq}$ at spin equilibrium for a given $M_\mathrm{acc}$, we must assume a range for $\Delta_\mathrm{eq}$, as the curve in Figure~\ref{fig:lognorm} is not monotonic. For $\Delta_\mathrm{eq} \gtrsim 0.04$, two values of $\Delta_\mathrm{eq}$ correspond to the same normalized spin equilibrium jet power, complicating inversion. Assuming that all lGRBs originate from systems where $\Delta_\mathrm{eq} < \Delta_\mathrm{eq,max}$ allows a one-to-one mapping between observed jet energy, $E_\mathrm{jet}$, and flux ratio $\Delta_\mathrm{eq}$. Figure~\ref{fig:phi_dis} (left) shows $\Delta_\mathrm{eq}$ distributions for various $M_\mathrm{acc}$ values relative to the observed jet energy distribution. As $M_\mathrm{acc}$ increases, the system deviates from the MAD regime, but even for $M_\mathrm{acc} = 0.2 M_\odot$, the peak of $\Delta_\mathrm{eq}$ remains below its maximum, aligning with our assumption that $\Delta_\mathrm{eq} < \Delta_\mathrm{eq,max}$ across all systems.

The MAD flux limit $\Phi_\mathrm{MAD,eq}(a_\mathrm{eq}, \dot{m}, M_\mathrm{BH})$ is primarily influenced by the mass accretion rate at a given $\Delta_\mathrm{eq}$, as $M_\mathrm{BH}$ remains almost constant during the burst, and $a_\mathrm{eq}$ depends solely on $\Delta_\mathrm{eq}$. We approximate $\dot{m} \approx M_\mathrm{acc}/T_{90}$, where $T_{90}$ is distributed as $\log_{10} T_{90} \sim \mathcal{N}(0.94, 0.464)$ \citep{2018Ap&SS.363..223Z}, based on \textit{Fermi} GRBs. Assuming this distribution is independent of jet luminosity, we derive the distribution of $\Phi_\mathrm{BH,eq}$, shown in Figure~\ref{fig:phi_dis} (right) shows the distributions of $\Phi_\mathrm{BH,eq}$ for $M_\mathrm{acc} = 0.2, 1, 5 M_\odot$. While the $\Delta_\mathrm{eq}$ distributions do not overlap, as $\dot{m}$ increases with $M_\mathrm{acc}$, the $\Phi_\mathrm{BH,eq}$ distributions overlap. This is because most systems are far from MAD, with $a_\mathrm{eq} \sim 1$, so the magnetic flux distributions closely follow the luminosity distribution. For $M_\mathrm{acc} = 0.2 M_\odot$, however, the high-flux tail should have $a_\mathrm{eq} < 1$ due to $\Delta_\mathrm{eq} > 0.1$, leading to a divergence from the other distributions. Our results remain largely unaffected when using alternative $T_{90}$ distributions, such as those derived from \textit{Swift}.

\begin{figure}
    \centering
    \includegraphics[width=\columnwidth]{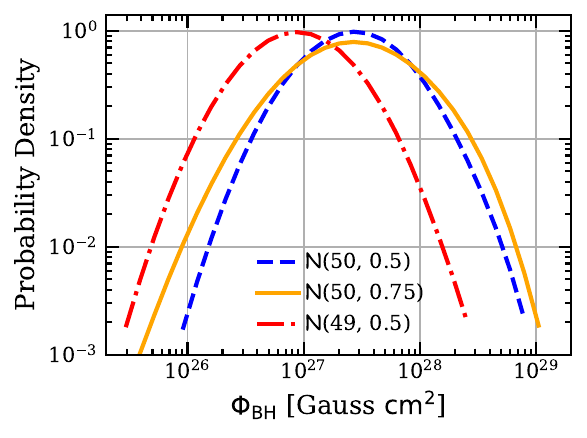}
    \caption{Probability distributions of magnetic flux threading the black hole, $\Phi_\mathrm{BH}$, for sGRBs. The blue dashed, orange solid, and red dash-dotted lines correspond to distributions of $\log_{10}(E_{\mathrm{jet}})$ following $\mathcal{N}(50, 0.5)$, $\mathcal{N}(50, 0.75)$, and $\mathcal{N}(49, 0.5)$, respectively. The $\Phi_\mathrm{BH}$ distributions for sGRBs resemble those of lGRBs and remain largely insensitive to reasonable variations in the $E_{\mathrm{jet}}$ distribution, owing to similar luminosity distributions and high black hole spin values during the bursts.}
    \label{fig:Phi_sGRBs}
\end{figure}

For sGRBs, assuming $E_{\mathrm{jet}} = L_{\mathrm{jet}}T_\mathrm{90}$, we find from Equation~\ref{L_BZ} that
\begin{equation}
    E_{\mathrm{jet}}=\frac{\kappa f}{4\pi c}\Phi_{\mathrm{BH}}^2 \Omega_\mathrm{H}^2(a=0.7, M_\mathrm{BH,0}) \,T_{90}
\end{equation}
Solving for $\Phi_{\mathrm{BH}}$ gives 
\begin{equation}
\begin{aligned}
      \Phi_{\mathrm{BH}}=2.77 &\times 10^{27}  \left(\frac{M_\mathrm{BH,0}}{2.5M_{\odot}}\right) \times \\
      & \left(\frac{E_{\text{jet}}}{10^{50}\text{erg}}\right)^{1/2}\left(\frac{T_{90}}{0.2\text{s}}\right)^{-1/2} [\text{Gauss }\text{cm}^2].
\end{aligned}
\end{equation}
From Equation~\ref{Phi_MAD}, the flux ratio $\Delta = \Phi_\mathrm{BH}/\Phi_\mathrm{MAD}$ becomes 
\begin{equation} \label{phiMADsGRb}
\begin{aligned}        
     \Delta & = \frac{\Phi_{\text{BH}}}{\Phi_{\text{MAD}}} \\
     & =0.028\left(\frac{E_{\text{jet}}} {10^{50}\text{erg}}\right)^{1/2}\left(\frac{T_{90}}{0.2\text{s}}\right)^{-1/2}\left(\frac{\dot{m}}{0.5 M_{\odot}/\text{s}}\right)^{-1/2}.
\end{aligned}
\end{equation}
For typical sGRB energies and durations, $\Phi_\mathrm{BH}$ is far from the MAD limit. Figure~\ref{fig:Phi_sGRBs} shows the distributions of $\Phi_\mathrm{BH}$ for sGRBs, assuming $\log_{10} T_{90} \sim \mathcal{N}(-0.62, 0.442)$ \citep{2018Ap&SS.363..223Z} and various distributions of $\log_{10}(E_{\mathrm{jet}})$. The $\Phi_\mathrm{BH}$ distributions for sGRBs closely resemble those of lGRBs due to similar luminosity distributions and high BH spin values during the bursts, and this similarity is largely insensitive to the specific choice of $E_\mathrm{jet}$ distribution, as long as it is reasonable. Again, we use the \textit{Fermi} GRB distribution for $\log_{10} T_{90}$, though the results are not sensitive to this specific choice. The magnetic flux distributions for both long and short GRBs peak around $10^{27} \text{ G cm}^2$, indicating a potential common origin of magnetic flux in both central engine systems.

\bibliography{Foison}{}

\begin{thebibliography}{}
\expandafter\ifx\csname natexlab\endcsname\relax\def\natexlab#1{#1}\fi
\providecommand{\url}[1]{\href{#1}{#1}}
\providecommand{\dodoi}[1]{doi:~\href{http://doi.org/#1}{\nolinkurl{#1}}}
\providecommand{\doeprint}[1]{\href{http://ascl.net/#1}{\nolinkurl{http://ascl.net/#1}}}
\providecommand{\doarXiv}[1]{\href{https://arxiv.org/abs/#1}{\nolinkurl{https://arxiv.org/abs/#1}}}

\bibitem[{{Abbott} {et~al.}(2017{\natexlab{a}}){Abbott}, {Abbott}, {Abbott}, {Acernese}, {Ackley}, {Adams}, {Adams}, {Addesso}, {Adhikari}, {Adya}, {Affeldt}, {Afrough}, {Agarwal}, {Agathos}, {Agatsuma}, {Aggarwal}, {Aguiar}, {Aiello}, {Ain}, {Ajith}, {Allen}, {Allen}, {Allocca}, {Altin}, {Amato}, {Ananyeva}, {Anderson}, {Anderson}, {Angelova}, {Antier}, {Appert}, {Arai}, {Araya}, {Areeda}, {Arnaud}, {Arun}, {Ascenzi}, {Ashton}, {Ast}, {Aston}, {Astone}, {Atallah}, {Aufmuth}, {Aulbert}, {AultONeal}, {Austin}, {Avila-Alvarez}, {Babak}, {Bacon}, {Bader}, {Bae}, {Bailes}, {Baker}, {Baldaccini}, {Ballardin}, {Ballmer}, {Banagiri}, {Barayoga}, {Barclay}, {Barish}, {Barker}, {Barkett}, {Barone}, {Barr}, {Barsotti}, {Barsuglia}, {Barta}, {Barthelmy}, {Bartlett}, {Bartos}, {Bassiri}, {Basti}, {Batch}, {Bawaj}, {Bayley}, {Bazzan}, {B{\'e}csy}, {Beer}, {Bejger}, {Belahcene}, {Bell}, {Berger}, {Bergmann}, {Bernuzzi}, {Bero}, {Berry}, {Bersanetti}, {Bertolini}, {Betzwieser}, {Bhagwat}, {Bhandare}, {Bilenko},
  {Billingsley}, {Billman}, {Birch}, {Birney}, {Birnholtz}, {Biscans}, {Biscoveanu}, {Bisht}, {Bitossi}, {Biwer}, {Bizouard}, {Blackburn}, {Blackman}, {Blair}, {Blair}, {Blair}, {Bloemen}, {Bock}, {Bode}, {Boer}, {Bogaert}, {Bohe}, {Bondu}, {Bonilla}, {Bonnand}, {Boom}, {Bork}, {Boschi}, {Bose}, {Bossie}, {Bouffanais}, {Bozzi}, {Bradaschia}, {Brady}, {Branchesi}, {Brau}, {Briant}, {Brillet}, {Brinkmann}, {Brisson}, {Brockill}, {Broida}, {Brooks}, {Brown}, {Brown}, {Brunett}, {Buchanan}, {Buikema}, {Bulik}, {Bulten}, {Buonanno}, {Buskulic}, {Buy}, {Byer}, {Cabero}, {Cadonati}, {Cagnoli}, {Cahillane}, {Calder{\'o}n Bustillo}, {Callister}, {Calloni}, {Camp}, {Canepa}, {Canizares}, {Cannon}, {Cao}, {Cao}, {Capano}, {Capocasa}, {Carbognani}, {Caride}, {Carney}, {Carullo}, {Casanueva Diaz}, {Casentini}, {Caudill}, {Cavagli{\`a}}, {Cavalier}, {Cavalieri}, {Cella}, {Cepeda}, {Cerd{\'a}-Dur{\'a}n}, {Cerretani}, {Cesarini}, {Chamberlin}, {Chan}, {Chao}, {Charlton}, {Chase}, {Chassande-Mottin}, {Chatterjee},
  {Chatziioannou}, {Cheeseboro}, {Chen}, {Chen}, {Chen}, {Cheng}, {Chia}, {Chincarini}, {Chiummo}, {Chmiel}, {Cho}, {Cho}, {Chow}, {Christensen}, {Chu}, {Chua}, {Chua}, {Chung}, {Chung}, {Ciani}, {Ciolfi}, {Cirelli}, {Cirone}, {Clara}, {Clark}, {Clearwater}, {Cleva}, {Cocchieri}, {Coccia}, {Cohadon}, {Cohen}, {Colla}, {Collette}, {Cominsky}, {Constancio}, {Conti}, {Cooper}, {Corban}, {Corbitt}, {Cordero-Carri{\'o}n}, {Corley}, {Cornish}, {Corsi}, {Cortese}, {Costa}, {Coughlin}, {Coughlin}, {Coulon}, {Countryman}, {Couvares}, {Covas}, {Cowan}, {Coward}, {Cowart}, {Coyne}, {Coyne}, {Creighton}, {Creighton}, {Cripe}, {Crowder}, {Cullen}, {Cumming}, {Cunningham}, {Cuoco}, {Dal Canton}, {D{\'a}lya}, {Danilishin}, {D'Antonio}, {Danzmann}, {Dasgupta}, {Da Silva Costa}, {Dattilo}, {Dave}, {Davier}, {Davis}, {Daw}, {Day}, {De}, {DeBra}, {Degallaix}, {De Laurentis}, {Del{\'e}glise}, {Del Pozzo}, {Demos}, {Denker}, {Dent}, {De Pietri}, {Dergachev}, {De Rosa}, {DeRosa}, {De Rossi}, {DeSalvo}, {de Varona}, {Devenson},
  {Dhurandhar}, {D{\'\i}az}, {Dietrich}, {Di Fiore}, {Di Giovanni}, {Di Girolamo}, {Di Lieto}, {Di Pace}, {Di Palma}, {Di Renzo}, {Doctor}, {Dolique}, {Donovan}, {Dooley}, {Doravari}, {Dorrington}, {Douglas}, {Dovale {\'A}lvarez}, {Downes}, {Drago}, {Dreissigacker}, {Driggers}, {Du}, {Ducrot}, {Dudi}, {Dupej}, {Dwyer}, {Edo}, {Edwards}, {Effler}, {Eggenstein}, {Ehrens}, {Eichholz}, {Eikenberry}, {Eisenstein}, {Essick}, {Estevez}, {Etienne}, {Etzel}, {Evans}, {Evans}, {Factourovich}, {Fafone}, {Fair}, {Fairhurst}, {Fan}, {Farinon}, {Farr}, {Farr}, {Fauchon-Jones}, {Favata}, {Fays}, {Fee}, {Fehrmann}, {Feicht}, {Fejer}, {Fernandez-Galiana}, {Ferrante}, {Ferreira}, {Ferrini}, {Fidecaro}, {Finstad}, {Fiori}, {Fiorucci}, {Fishbach}, {Fisher}, {Fitz-Axen}, {Flaminio}, {Fletcher}, {Fong}, {Font}, {Forsyth}, {Forsyth}, {Fournier}, {Frasca}, {Frasconi}, {Frei}, {Freise}, {Frey}, {Frey}, {Fries}, {Fritschel}, {Frolov}, {Fulda}, {Fyffe}, {Gabbard}, {Gadre}, {Gaebel}, {Gair}, {Gammaitoni}, {Ganija}, {Gaonkar},
  {Garcia-Quiros}, {Garufi}, {Gateley}, {Gaudio}, {Gaur}, {Gayathri}, {Gehrels}, {Gemme}, {Genin}, {Gennai}, {George}, {George}, {Gergely}, {Germain}, {Ghonge}, {Ghosh}, {Ghosh}, {Ghosh}, {Giaime}, {Giardina}, {Giazotto}, {Gill}, {Glover}, {Goetz}, {Goetz}, {Gomes}, {Goncharov}, {Gonz{\'a}lez}, {Gonzalez Castro}, {Gopakumar}, {Gorodetsky}, {Gossan}, {Gosselin}, {Gouaty}, {Grado}, {Graef}, {Granata}, {Grant}, {Gras}, {Gray}, {Greco}, {Green}, {Gretarsson}, {Groot}, {Grote}, {Grunewald}, {Gruning}, {Guidi}, {Guo}, {Gupta}, {Gupta}, {Gushwa}, {Gustafson}, {Gustafson}, {Halim}, {Hall}, {Hall}, {Hamilton}, {Hammond}, {Haney}, {Hanke}, {Hanks}, {Hanna}, {Hannam}, {Hannuksela}, {Hanson}, {Hardwick}, {Harms}, {Harry}, {Harry}, {Hart}, {Haster}, {Haughian}, {Healy}, {Heidmann}, {Heintze}, {Heitmann}, {Hello}, {Hemming}, {Hendry}, {Heng}, {Hennig}, {Heptonstall}, {Heurs}, {Hild}, {Hinderer}, {Ho}, {Hoak}, {Hofman}, {Holt}, {Holz}, {Hopkins}, {Horst}, {Hough}, {Houston}, {Howell}, {Hreibi}, {Hu}, {Huerta}, {Huet},
  {Hughey}, {Husa}, {Huttner}, {Huynh-Dinh}, {Indik}, {Inta}, {Intini}, {Isa}, {Isac}, {Isi}, {Iyer}, {Izumi}, {Jacqmin}, {Jani}, {Jaranowski}, {Jawahar}, {Jim{\'e}nez-Forteza}, {Johnson}, {Johnson-McDaniel}, {Jones}, {Jones}, {Jonker}, {Ju}, {Junker}, {Kalaghatgi}, {Kalogera}, {Kamai}, {Kandhasamy}, {Kang}, {Kanner}, {Kapadia}, {Karki}, {Karvinen}, {Kasprzack}, {Kastaun}, {Katolik}, {Katsavounidis}, {Katzman}, {Kaufer}, {Kawabe}, {K{\'e}f{\'e}lian}, {Keitel}, {Kemball}, {Kennedy}, {Kent}, {Key}, {Khalili}, {Khan}, {Khan}, {Khan}, {Khazanov}, {Kijbunchoo}, {Kim}, {Kim}, {Kim}, {Kim}, {Kim}, {Kim}, {Kimbrell}, {King}, {King}, {Kinley-Hanlon}, {Kirchhoff}, {Kissel}, {Kleybolte}, {Klimenko}, {Knowles}, {Koch}, {Koehlenbeck}, {Koley}, {Kondrashov}, {Kontos}, {Korobko}, {Korth}, {Kowalska}, {Kozak}, {Kr{\"a}mer}, {Kringel}, {Krishnan}, {Kr{\'o}lak}, {Kuehn}, {Kumar}, {Kumar}, {Kumar}, {Kuo}, {Kutynia}, {Kwang}, {Lackey}, {Lai}, {Landry}, {Lang}, {Lange}, {Lantz}, {Lanza}, {Larson}, {Lartaux-Vollard}, {Lasky},
  {Laxen}, {Lazzarini}, {Lazzaro}, {Leaci}, {Leavey}, {Lee}, {Lee}, {Lee}, {Lee}, {Lee}, {Lehmann}, {Lenon}, {Leon}, {Leonardi}, {Leroy}, {Letendre}, {Levin}, {Li}, {Linker}, {Littenberg}, {Liu}, {Liu}, {Lo}, {Lockerbie}, {London}, {Lord}, {Lorenzini}, {Loriette}, {Lormand}, {Losurdo}, {Lough}, {Lousto}, {Lovelace}, {L{\"u}ck}, {Lumaca}, {Lundgren}, {Lynch}, {Ma}, {Macas}, {Macfoy}, {Machenschalk}, {MacInnis}, {Macleod}, {Maga{\~n}a Hernandez}, {Maga{\~n}a-Sandoval}, {Maga{\~n}a Zertuche}, {Magee}, {Majorana}, {Maksimovic}, {Man}, {Mandic}, {Mangano}, {Mansell}, {Manske}, {Mantovani}, {Marchesoni}, {Marion}, {M{\'a}rka}, {M{\'a}rka}, {Markakis}, {Markosyan}, {Markowitz}, {Maros}, {Marquina}, {Marsh}, {Martelli}, {Martellini}, {Martin}, {Martin}, {Martynov}, {Marx}, {Mason}, {Massera}, {Masserot}, {Massinger}, {Masso-Reid}, {Mastrogiovanni}, {Matas}, {Matichard}, {Matone}, {Mavalvala}, {Mazumder}, {McCarthy}, {McClelland}, {McCormick}, {McCuller}, {McGuire}, {McIntyre}, {McIver}, {McManus}, {McNeill}, {McRae},
  {McWilliams}, {Meacher}, {Meadors}, {Mehmet}, {Meidam}, {Mejuto-Villa}, {Melatos}, {Mendell}, {Mercer}, {Merilh}, {Merzougui}, {Meshkov}, {Messenger}, {Messick}, {Metzdorff}, {Meyers}, {Miao}, {Michel}, {Middleton}, {Mikhailov}, {Milano}, {Miller}, {Miller}, {Miller}, {Millhouse}, {Milovich-Goff}, {Minazzoli}, {Minenkov}, {Ming}, {Mishra}, {Mitra}, {Mitrofanov}, {Mitselmakher}, {Mittleman}, {Moffa}, {Moggi}, {Mogushi}, {Mohan}, {Mohapatra}, {Molina}, {Montani}, {Moore}, {Moraru}, {Moreno}, {Morisaki}, {Morriss}, {Mours}, {Mow-Lowry}, {Mueller}, {Muir}, {Mukherjee}, {Mukherjee}, {Mukherjee}, {Mukund}, {Mullavey}, {Munch}, {Mu{\~n}iz}, {Muratore}, {Murray}, {Nagar}, {Napier}, {Nardecchia}, {Naticchioni}, {Nayak}, {Neilson}, {Nelemans}, {Nelson}, {Nery}, {Neunzert}, {Nevin}, {Newport}, {Newton}, {Ng}, {Nguyen}, {Nguyen}, {Nichols}, {Nielsen}, {Nissanke}, {Nitz}, {Noack}, {Nocera}, {Nolting}, {North}, {Nuttall}, {Oberling}, {O'Dea}, {Ogin}, {Oh}, {Oh}, {Ohme}, {Okada}, {Oliver}, {Oppermann}, {Oram}, {O'Reilly},
  {Ormiston}, {Ortega}, {O'Shaughnessy}, {Ossokine}, {Ottaway}, {Overmier}, {Owen}, {Pace}, {Page}, {Page}, {Pai}, {Pai}, {Palamos}, {Palashov}, {Palomba}, {Pal-Singh}, {Pan}, {Pan}, {Pang}, {Pang}, {Pankow}, {Pannarale}, {Pant}, {Paoletti}, {Paoli}, {Papa}, {Parida}, {Parker}, {Pascucci}, {Pasqualetti}, {Passaquieti}, {Passuello}, {Patil}, {Patricelli}, {Pearlstone}, {Pedraza}, {Pedurand}, {Pekowsky}, {Pele}, {Penn}, {Perez}, {Perreca}, {Perri}, {Pfeiffer}, {Phelps}, {Piccinni}, {Pichot}, {Piergiovanni}, {Pierro}, {Pillant}, {Pinard}, {Pinto}, {Pirello}, {Pitkin}, {Poe}, {Poggiani}, {Popolizio}, {Porter}, {Post}, {Powell}, {Prasad}, {Pratt}, {Pratten}, {Predoi}, {Prestegard}, {Prijatelj}, {Principe}, {Privitera}, {Prix}, {Prodi}, {Prokhorov}, {Puncken}, {Punturo}, {Puppo}, {P{\"u}rrer}, {Qi}, {Quetschke}, {Quintero}, {Quitzow-James}, {Raab}, {Rabeling}, {Radkins}, {Raffai}, {Raja}, {Rajan}, {Rajbhandari}, {Rakhmanov}, {Ramirez}, {Ramos-Buades}, {Rapagnani}, {Raymond}, {Razzano}, {Read}, {Regimbau}, {Rei},
  {Reid}, {Reitze}, {Ren}, {Reyes}, {Ricci}, {Ricker}, {Rieger}, {Riles}, {Rizzo}, {Robertson}, {Robie}, {Robinet}, {Rocchi}, {Rolland}, {Rollins}, {Roma}, {Romano}, {Romano}, {Romel}, {Romie}, {Rosi{\'n}ska}, {Ross}, {Rowan}, {R{\"u}diger}, {Ruggi}, {Rutins}, {Ryan}, {Sachdev}, {Sadecki}, {Sadeghian}, {Sakellariadou}, {Salconi}, {Saleem}, {Salemi}, {Samajdar}, {Sammut}, {Sampson}, {Sanchez}, {Sanchez}, {Sanchis-Gual}, {Sandberg}, {Sanders}, {Sassolas}, {Sathyaprakash}, {Saulson}, {Sauter}, {Savage}, {Sawadsky}, {Schale}, {Scheel}, {Scheuer}, {Schmidt}, {Schmidt}, {Schnabel}, {Schofield}, {Sch{\"o}nbeck}, {Schreiber}, {Schuette}, {Schulte}, {Schutz}, {Schwalbe}, {Scott}, {Scott}, {Seidel}, {Sellers}, {Sengupta}, {Sentenac}, {Sequino}, {Sergeev}, {Shaddock}, {Shaffer}, {Shah}, {Shahriar}, {Shaner}, {Shao}, {Shapiro}, {Shawhan}, {Sheperd}, {Shoemaker}, {Shoemaker}, {Siellez}, {Siemens}, {Sieniawska}, {Sigg}, {Silva}, {Singer}, {Singh}, {Singhal}, {Sintes}, {Slagmolen}, {Smith}, {Smith}, {Smith}, {Somala},
  {Son}, {Sonnenberg}, {Sorazu}, {Sorrentino}, {Souradeep}, {Spencer}, {Srivastava}, {Staats}, {Staley}, {Steinke}, {Steinlechner}, {Steinlechner}, {Steinmeyer}, {Stevenson}, {Stone}, {Stops}, {Strain}, {Stratta}, {Strigin}, {Strunk}, {Sturani}, {Stuver}, {Summerscales}, {Sun}, {Sunil}, {Suresh}, {Sutton}, {Swinkels}, {Szczepa{\'n}czyk}, {Tacca}, {Tait}, {Talbot}, {Talukder}, {Tanner}, {T{\'a}pai}, {Taracchini}, {Tasson}, {Taylor}, {Taylor}, {Tewari}, {Theeg}, {Thies}, {Thomas}, {Thomas}, {Thomas}, {Thorne}, {Thorne}, {Thrane}, {Tiwari}, {Tiwari}, {Tokmakov}, {Toland}, {Tonelli}, {Tornasi}, {Torres-Forn{\'e}}, {Torrie}, {T{\"o}yr{\"a}}, {Travasso}, {Traylor}, {Trinastic}, {Tringali}, {Trozzo}, {Tsang}, {Tse}, {Tso}, {Tsukada}, {Tsuna}, {Tuyenbayev}, {Ueno}, {Ugolini}, {Unnikrishnan}, {Urban}, {Usman}, {Vahlbruch}, {Vajente}, {Valdes}, {Vallisneri}, {van Bakel}, {van Beuzekom}, {van den Brand}, {Van Den Broeck}, {Vander-Hyde}, {van der Schaaf}, {van Heijningen}, {van Veggel}, {Vardaro}, {Varma}, {Vass},
  {Vas{\'u}th}, {Vecchio}, {Vedovato}, {Veitch}, {Veitch}, {Venkateswara}, {Venugopalan}, {Verkindt}, {Vetrano}, {Vicer{\'e}}, {Viets}, {Vinciguerra}, {Vine}, {Vinet}, {Vitale}, {Vo}, {Vocca}, {Vorvick}, {Vyatchanin}, {Wade}, {Wade}, {Wade}, {Walet}, {Walker}, {Wallace}, {Walsh}, {Wang}, {Wang}, {Wang}, {Wang}, {Wang}, {Ward}, {Warner}, {Was}, {Watchi}, {Weaver}, {Wei}, {Weinert}, {Weinstein}, {Weiss}, {Wen}, {Wessel}, {We{\ss}els}, {Westerweck}, {Westphal}, {Wette}, {Whelan}, {Whitcomb}, {Whiting}, {Whittle}, {Wilken}, {Williams}, {Williams}, {Williamson}, {Willis}, {Willke}, {Wimmer}, {Winkler}, {Wipf}, {Wittel}, {Woan}, {Woehler}, {Wofford}, {Wong}, {Worden}, {Wright}, {Wu}, {Wysocki}, {Xiao}, {Yamamoto}, {Yancey}, {Yang}, {Yap}, {Yazback}, {Yu}, {Yu}, {Yvert}, {Zadro{\.Z}ny}, {Zanolin}, {Zelenova}, {Zendri}, {Zevin}, {Zhang}, {Zhang}, {Zhang}, {Zhang}, {Zhao}, {Zhou}, {Zhou}, {Zhu}, {Zhu}, {Zimmerman}, {Zucker}, {Zweizig}, {LIGO Scientific Collaboration}, \& {Virgo Collaboration}}]{2017PhRvL.119p1101A}
{Abbott}, B.~P., {Abbott}, R., {Abbott}, T.~D., {et~al.} 2017{\natexlab{a}}, \prl, 119, 161101, \dodoi{10.1103/PhysRevLett.119.161101}

\bibitem[{{Abbott} {et~al.}(2017{\natexlab{b}}){Abbott}, {Abbott}, {Abbott}, {Acernese}, {Ackley}, {Adams}, {Adams}, {Addesso}, {Adhikari}, {Adya}, {Affeldt}, {Afrough}, {Agarwal}, {Agathos}, {Agatsuma}, {Aggarwal}, {Aguiar}, {Aiello}, {Ain}, {Ajith}, {Allen}, {Allen}, {Allocca}, {Aloy}, {Altin}, {Amato}, {Ananyeva}, {Anderson}, {Anderson}, {Angelova}, {Antier}, {Appert}, {Arai}, {Araya}, {Areeda}, {Arnaud}, {Arun}, {Ascenzi}, {Ashton}, {Ast}, {Aston}, {Astone}, {Atallah}, {Aufmuth}, {Aulbert}, {AultONeal}, {Austin}, {Avila-Alvarez}, {Babak}, {Bacon}, {Bader}, {Bae}, {Baker}, {Baldaccini}, {Ballardin}, {Ballmer}, {Banagiri}, {Barayoga}, {Barclay}, {Barish}, {Barker}, {Barkett}, {Barone}, {Barr}, {Barsotti}, {Barsuglia}, {Barta}, {Bartlett}, {Bartos}, {Bassiri}, {Basti}, {Batch}, {Bawaj}, {Bayley}, {Bazzan}, {B{\'e}csy}, {Beer}, {Bejger}, {Belahcene}, {Bell}, {Berger}, {Bergmann}, {Bero}, {Berry}, {Bersanetti}, {Bertolini}, {Betzwieser}, {Bhagwat}, {Bhandare}, {Bilenko}, {Billingsley}, {Billman}, {Birch},
  {Birney}, {Birnholtz}, {Biscans}, {Biscoveanu}, {Bisht}, {Bitossi}, {Biwer}, {Bizouard}, {Blackburn}, {Blackman}, {Blair}, {Blair}, {Blair}, {Bloemen}, {Bock}, {Bode}, {Boer}, {Bogaert}, {Bohe}, {Bondu}, {Bonilla}, {Bonnand}, {Boom}, {Bork}, {Boschi}, {Bose}, {Bossie}, {Bouffanais}, {Bozzi}, {Bradaschia}, {Brady}, {Branchesi}, {Brau}, {Briant}, {Brillet}, {Brinkmann}, {Brisson}, {Brockill}, {Broida}, {Brooks}, {Brown}, {Brown}, {Brunett}, {Buchanan}, {Buikema}, {Bulik}, {Bulten}, {Buonanno}, {Buskulic}, {Buy}, {Byer}, {Cabero}, {Cadonati}, {Cagnoli}, {Cahillane}, {Calder{\'o}n Bustillo}, {Callister}, {Calloni}, {Camp}, {Canepa}, {Canizares}, {Cannon}, {Cao}, {Cao}, {Capano}, {Capocasa}, {Carbognani}, {Caride}, {Carney}, {Casanueva Diaz}, {Casentini}, {Caudill}, {Cavagli{\`a}}, {Cavalier}, {Cavalieri}, {Cella}, {Cepeda}, {Cerd{\'a}-Dur{\'a}n}, {Cerretani}, {Cesarini}, {Chamberlin}, {Chan}, {Chao}, {Charlton}, {Chase}, {Chassande-Mottin}, {Chatterjee}, {Chatziioannou}, {Cheeseboro}, {Chen}, {Chen}, {Chen},
  {Cheng}, {Chia}, {Chincarini}, {Chiummo}, {Chmiel}, {Cho}, {Cho}, {Chow}, {Christensen}, {Chu}, {Chua}, {Chua}, {Chung}, {Chung}, {Ciani}, {Ciolfi}, {Cirelli}, {Cirone}, {Clara}, {Clark}, {Clearwater}, {Cleva}, {Cocchieri}, {Coccia}, {Cohadon}, {Cohen}, {Colla}, {Collette}, {Cominsky}, {Constancio}, {Conti}, {Cooper}, {Corban}, {Corbitt}, {Cordero-Carri{\'o}n}, {Corley}, {Cornish}, {Corsi}, {Cortese}, {Costa}, {Coughlin}, {Coughlin}, {Coulon}, {Countryman}, {Couvares}, {Covas}, {Cowan}, {Coward}, {Cowart}, {Coyne}, {Coyne}, {Creighton}, {Creighton}, {Cripe}, {Crowder}, {Cullen}, {Cumming}, {Cunningham}, {Cuoco}, {Dal Canton}, {D{\'a}lya}, {Danilishin}, {D'Antonio}, {Danzmann}, {Dasgupta}, {Da Silva Costa}, {Dattilo}, {Dave}, {Davier}, {Davis}, {Daw}, {Day}, {De}, {DeBra}, {Degallaix}, {De Laurentis}, {Del{\'e}glise}, {Del Pozzo}, {Demos}, {Denker}, {Dent}, {De Pietri}, {Dergachev}, {De Rosa}, {DeRosa}, {De Rossi}, {DeSalvo}, {de Varona}, {Devenson}, {Dhurandhar}, {D{\'\i}az}, {Di Fiore}, {Di Giovanni}, {Di
  Girolamo}, {Di Lieto}, {Di Pace}, {Di Palma}, {Di Renzo}, {Doctor}, {Dolique}, {Donovan}, {Dooley}, {Doravari}, {Dorrington}, {Douglas}, {Dovale {\'A}lvarez}, {Downes}, {Drago}, {Dreissigacker}, {Driggers}, {Du}, {Ducrot}, {Dupej}, {Dwyer}, {Edo}, {Edwards}, {Effler}, {Eggenstein}, {Ehrens}, {Eichholz}, {Eikenberry}, {Eisenstein}, {Essick}, {Estevez}, {Etienne}, {Etzel}, {Evans}, {Evans}, {Factourovich}, {Fafone}, {Fair}, {Fairhurst}, {Fan}, {Farinon}, {Farr}, {Farr}, {Fauchon-Jones}, {Favata}, {Fays}, {Fee}, {Fehrmann}, {Feicht}, {Fejer}, {Fernandez-Galiana}, {Ferrante}, {Ferreira}, {Ferrini}, {Fidecaro}, {Finstad}, {Fiori}, {Fiorucci}, {Fishbach}, {Fisher}, {Fitz-Axen}, {Flaminio}, {Fletcher}, {Fong}, {Font}, {Forsyth}, {Forsyth}, {Fournier}, {Frasca}, {Frasconi}, {Frei}, {Freise}, {Frey}, {Frey}, {Fries}, {Fritschel}, {Frolov}, {Fulda}, {Fyffe}, {Gabbard}, {Gadre}, {Gaebel}, {Gair}, {Gammaitoni}, {Ganija}, {Gaonkar}, {Garcia-Quiros}, {Garufi}, {Gateley}, {Gaudio}, {Gaur}, {Gayathri}, {Gehrels}, {Gemme},
  {Genin}, {Gennai}, {George}, {George}, {Gergely}, {Germain}, {Ghonge}, {Ghosh}, {Ghosh}, {Ghosh}, {Giaime}, {Giardina}, {Giazotto}, {Gill}, {Glover}, {Goetz}, {Goetz}, {Gomes}, {Goncharov}, {Gonz{\'a}lez}, {Gonzalez Castro}, {Gopakumar}, {Gorodetsky}, {Gossan}, {Gosselin}, {Gouaty}, {Grado}, {Graef}, {Granata}, {Grant}, {Gras}, {Gray}, {Greco}, {Green}, {Gretarsson}, {Groot}, {Grote}, {Grunewald}, {Gruning}, {Guidi}, {Guo}, {Gupta}, {Gupta}, {Gushwa}, {Gustafson}, {Gustafson}, {Halim}, {Hall}, {Hall}, {Hamilton}, {Hammond}, {Haney}, {Hanke}, {Hanks}, {Hanna}, {Hannam}, {Hannuksela}, {Hanson}, {Hardwick}, {Harms}, {Harry}, {Harry}, {Hart}, {Haster}, {Haughian}, {Healy}, {Heidmann}, {Heintze}, {Heitmann}, {Hello}, {Hemming}, {Hendry}, {Heng}, {Hennig}, {Heptonstall}, {Heurs}, {Hild}, {Hinderer}, {Hoak}, {Hofman}, {Holt}, {Holz}, {Hopkins}, {Horst}, {Hough}, {Houston}, {Howell}, {Hreibi}, {Hu}, {Huerta}, {Huet}, {Hughey}, {Husa}, {Huttner}, {Huynh-Dinh}, {Indik}, {Inta}, {Intini}, {Isa}, {Isac}, {Isi}, {Iyer},
  {Izumi}, {Jacqmin}, {Jani}, {Jaranowski}, {Jawahar}, {Jim{\'e}nez-Forteza}, {Johnson}, {Johnson-McDaniel}, {Jones}, {Jones}, {Jonker}, {Ju}, {Junker}, {Kalaghatgi}, {Kalogera}, {Kamai}, {Kandhasamy}, {Kang}, {Kanner}, {Kapadia}, {Karki}, {Karvinen}, {Kasprzack}, {Kastaun}, {Katolik}, {Katsavounidis}, {Katzman}, {Kaufer}, {Kawabe}, {K{\'e}f{\'e}lian}, {Keitel}, {Kemball}, {Kennedy}, {Kent}, {Key}, {Khalili}, {Khan}, {Khan}, {Khan}, {Khazanov}, {Kijbunchoo}, {Kim}, {Kim}, {Kim}, {Kim}, {Kim}, {Kim}, {Kimbrell}, {King}, {King}, {Kinley-Hanlon}, {Kirchhoff}, {Kissel}, {Kleybolte}, {Klimenko}, {Knowles}, {Koch}, {Koehlenbeck}, {Koley}, {Kondrashov}, {Kontos}, {Korobko}, {Korth}, {Kowalska}, {Kozak}, {Kr{\"a}mer}, {Kringel}, {Krishnan}, {Kr{\'o}lak}, {Kuehn}, {Kumar}, {Kumar}, {Kumar}, {Kuo}, {Kutynia}, {Kwang}, {Lackey}, {Lai}, {Landry}, {Lang}, {Lange}, {Lantz}, {Lanza}, {Lartaux-Vollard}, {Lasky}, {Laxen}, {Lazzarini}, {Lazzaro}, {Leaci}, {Leavey}, {Lee}, {Lee}, {Lee}, {Lee}, {Lee}, {Lehmann}, {Lenon},
  {Leonardi}, {Leroy}, {Letendre}, {Levin}, {Li}, {Linker}, {Littenberg}, {Liu}, {Lo}, {Lockerbie}, {London}, {Lord}, {Lorenzini}, {Loriette}, {Lormand}, {Losurdo}, {Lough}, {Lousto}, {Lovelace}, {L{\"u}ck}, {Lumaca}, {Lundgren}, {Lynch}, {Ma}, {Macas}, {Macfoy}, {Machenschalk}, {MacInnis}, {Macleod}, {Maga{\~n}a Hernandez}, {Maga{\~n}a-Sandoval}, {Maga{\~n}a Zertuche}, {Magee}, {Majorana}, {Maksimovic}, {Man}, {Mandic}, {Mangano}, {Mansell}, {Manske}, {Mantovani}, {Marchesoni}, {Marion}, {M{\'a}rka}, {M{\'a}rka}, {Markakis}, {Markosyan}, {Markowitz}, {Maros}, {Marquina}, {Martelli}, {Martellini}, {Martin}, {Martin}, {Martynov}, {Mason}, {Massera}, {Masserot}, {Massinger}, {Masso-Reid}, {Mastrogiovanni}, {Matas}, {Matichard}, {Matone}, {Mavalvala}, {Mazumder}, {McCarthy}, {McClelland}, {McCormick}, {McCuller}, {McGuire}, {McIntyre}, {McIver}, {McManus}, {McNeill}, {McRae}, {McWilliams}, {Meacher}, {Meadors}, {Mehmet}, {Meidam}, {Mejuto-Villa}, {Melatos}, {Mendell}, {Mercer}, {Merilh}, {Merzougui}, {Meshkov},
  {Messenger}, {Messick}, {Metzdorff}, {Meyers}, {Miao}, {Michel}, {Middleton}, {Mikhailov}, {Milano}, {Miller}, {Miller}, {Miller}, {Millhouse}, {Milovich-Goff}, {Minazzoli}, {Minenkov}, {Ming}, {Mishra}, {Mitra}, {Mitrofanov}, {Mitselmakher}, {Mittleman}, {Moffa}, {Moggi}, {Mogushi}, {Mohan}, {Mohapatra}, {Montani}, {Moore}, {Moraru}, {Moreno}, {Morriss}, {Mours}, {Mow-Lowry}, {Mueller}, {Muir}, {Mukherjee}, {Mukherjee}, {Mukherjee}, {Mukund}, {Mullavey}, {Munch}, {Mu{\~n}iz}, {Muratore}, {Murray}, {Napier}, {Nardecchia}, {Naticchioni}, {Nayak}, {Neilson}, {Nelemans}, {Nelson}, {Nery}, {Neunzert}, {Nevin}, {Newport}, {Newton}, {Ng}, {Nguyen}, {Nichols}, {Nielsen}, {Nissanke}, {Nitz}, {Noack}, {Nocera}, {Nolting}, {North}, {Nuttall}, {Oberling}, {O'Dea}, {Ogin}, {Oh}, {Oh}, {Ohme}, {Okada}, {Oliver}, {Oppermann}, {Oram}, {O'Reilly}, {Ormiston}, {Ortega}, {O'Shaughnessy}, {Ossokine}, {Ottaway}, {Overmier}, {Owen}, {Pace}, {Page}, {Page}, {Pai}, {Pai}, {Palamos}, {Palashov}, {Palomba}, {Pal-Singh}, {Pan},
  {Pan}, {Pang}, {Pang}, {Pankow}, {Pannarale}, {Pant}, {Paoletti}, {Paoli}, {Papa}, {Parida}, {Parker}, {Pascucci}, {Pasqualetti}, {Passaquieti}, {Passuello}, {Patil}, {Patricelli}, {Pearlstone}, {Pedraza}, {Pedurand}, {Pekowsky}, {Pele}, {Penn}, {Perez}, {Perreca}, {Perri}, {Pfeiffer}, {Phelps}, {Piccinni}, {Pichot}, {Piergiovanni}, {Pierro}, {Pillant}, {Pinard}, {Pinto}, {Pirello}, {Pitkin}, {Poe}, {Poggiani}, {Popolizio}, {Porter}, {Post}, {Powell}, {Prasad}, {Pratt}, {Pratten}, {Predoi}, {Prestegard}, {Prijatelj}, {Principe}, {Privitera}, {Prodi}, {Prokhorov}, {Puncken}, {Punturo}, {Puppo}, {P{\"u}rrer}, {Qi}, {Quetschke}, {Quintero}, {Quitzow-James}, {Raab}, {Rabeling}, {Radkins}, {Raffai}, {Raja}, {Rajan}, {Rajbhandari}, {Rakhmanov}, {Ramirez}, {Ramos-Buades}, {Rapagnani}, {Raymond}, {Razzano}, {Read}, {Regimbau}, {Rei}, {Reid}, {Reitze}, {Ren}, {Reyes}, {Ricci}, {Ricker}, {Rieger}, {Riles}, {Rizzo}, {Robertson}, {Robie}, {Robinet}, {Rocchi}, {Rolland}, {Rollins}, {Roma}, {Romano}, {Romel}, {Romie},
  {Rosi{\'n}ska}, {Ross}, {Rowan}, {R{\"u}diger}, {Ruggi}, {Rutins}, {Ryan}, {Sachdev}, {Sadecki}, {Sadeghian}, {Sakellariadou}, {Salconi}, {Saleem}, {Salemi}, {Samajdar}, {Sammut}, {Sampson}, {Sanchez}, {Sanchez}, {Sanchis-Gual}, {Sandberg}, {Sanders}, {Sassolas}, {Sathyaprakash}, {Saulson}, {Sauter}, {Savage}, {Sawadsky}, {Schale}, {Scheel}, {Scheuer}, {Schmidt}, {Schmidt}, {Schnabel}, {Schofield}, {Sch{\"o}nbeck}, {Schreiber}, {Schuette}, {Schulte}, {Schutz}, {Schwalbe}, {Scott}, {Scott}, {Seidel}, {Sellers}, {Sengupta}, {Sentenac}, {Sequino}, {Sergeev}, {Shaddock}, {Shaffer}, {Shah}, {Shahriar}, {Shaner}, {Shao}, {Shapiro}, {Shawhan}, {Sheperd}, {Shoemaker}, {Shoemaker}, {Siellez}, {Siemens}, {Sieniawska}, {Sigg}, {Silva}, {Singer}, {Singh}, {Singhal}, {Sintes}, {Slagmolen}, {Smith}, {Smith}, {Smith}, {Somala}, {Son}, {Sonnenberg}, {Sorazu}, {Sorrentino}, {Souradeep}, {Spencer}, {Srivastava}, {Staats}, {Staley}, {Steinke}, {Steinlechner}, {Steinlechner}, {Steinmeyer}, {Stevenson}, {Stone}, {Stops},
  {Strain}, {Stratta}, {Strigin}, {Strunk}, {Sturani}, {Stuver}, {Summerscales}, {Sun}, {Sunil}, {Suresh}, {Sutton}, {Swinkels}, {Szczepa{\'n}czyk}, {Tacca}, {Tait}, {Talbot}, {Talukder}, {Tanner}, {T{\'a}pai}, {Taracchini}, {Tasson}, {Taylor}, {Taylor}, {Tewari}, {Theeg}, {Thies}, {Thomas}, {Thomas}, {Thomas}, {Thorne}, {Thorne}, {Thrane}, {Tiwari}, {Tiwari}, {Tokmakov}, {Toland}, {Tonelli}, {Tornasi}, {Torres-Forn{\'e}}, {Torrie}, {T{\"o}yr{\"a}}, {Travasso}, {Traylor}, {Trinastic}, {Tringali}, {Trozzo}, {Tsang}, {Tse}, {Tso}, {Tsukada}, {Tsuna}, {Tuyenbayev}, {Ueno}, {Ugolini}, {Unnikrishnan}, {Urban}, {Usman}, {Vahlbruch}, {Vajente}, {Valdes}, {van Bakel}, {van Beuzekom}, {van den Brand}, {Van Den Broeck}, {Vander-Hyde}, {van der Schaaf}, {van Heijningen}, {van Veggel}, {Vardaro}, {Varma}, {Vass}, {Vas{\'u}th}, {Vecchio}, {Vedovato}, {Veitch}, {Veitch}, {Venkateswara}, {Venugopalan}, {Verkindt}, {Vetrano}, {Vicer{\'e}}, {Viets}, {Vinciguerra}, {Vine}, {Vinet}, {Vitale}, {Vo}, {Vocca}, {Vorvick},
  {Vyatchanin}, {Wade}, {Wade}, {Wade}, {Walet}, {Walker}, {Wallace}, {Walsh}, {Wang}, {Wang}, {Wang}, {Wang}, {Wang}, {Ward}, {Warner}, {Was}, {Watchi}, {Weaver}, {Wei}, {Weinert}, {Weinstein}, {Weiss}, {Wen}, {Wessel}, {We{\ss}els}, {Westerweck}, {Westphal}, {Wette}, {Whelan}, {Whitcomb}, {Whiting}, {Whittle}, {Wilken}, {Williams}, {Williams}, {Williamson}, {Willis}, {Willke}, {Wimmer}, {Winkler}, {Wipf}, {Wittel}, {Woan}, {Woehler}, {Wofford}, {Wong}, {Worden}, {Wright}, {Wu}, {Wysocki}, {Xiao}, {Yamamoto}, {Yancey}, {Yang}, {Yap}, {Yazback}, {Yu}, {Yu}, {Yvert}, {Zadro{\.z}ny}, {Zanolin}, {Zelenova}, {Zendri}, {Zevin}, {Zhang}, {Zhang}, {Zhang}, {Zhang}, {Zhao}, {Zhou}, {Zhou}, {Zhu}, {Zhu}, {Zimmerman}, {Zucker}, {Zweizig}, {(LIGO Scientific Collaboration}, {Virgo Collaboration}, {Burns}, {Veres}, {Kocevski}, {Racusin}, {Goldstein}, {Connaughton}, {Briggs}, {Blackburn}, {Hamburg}, {Hui}, {von Kienlin}, {McEnery}, {Preece}, {Wilson-Hodge}, {Bissaldi}, {Cleveland}, {Gibby}, {Giles}, {Kippen}, {McBreen},
  {Meegan}, {Paciesas}, {Poolakkil}, {Roberts}, {Stanbro}, {Gamma-ray Burst Monitor}, {Savchenko}, {Ferrigno}, {Kuulkers}, {Bazzano}, {Bozzo}, {Brandt}, {Chenevez}, {Courvoisier}, {Diehl}, {Domingo}, {Hanlon}, {Jourdain}, {Laurent}, {Lebrun}, {Lutovinov}, {Mereghetti}, {Natalucci}, {Rodi}, {Roques}, {Sunyaev}, {Ubertini}, \& {(INTEGRAL}}]{2017ApJ...848L..13A}
---. 2017{\natexlab{b}}, \apjl, 848, L13, \dodoi{10.3847/2041-8213/aa920c}

\bibitem[{{Bardeen}(1970)}]{1970Natur.226...64B}
{Bardeen}, J.~M. 1970, \nat, 226, 64, \dodoi{10.1038/226064a0}

\bibitem[{{Beniamini} {et~al.}(2017){Beniamini}, {Giannios}, \& {Metzger}}]{2017MNRAS.472.3058B}
{Beniamini}, P., {Giannios}, D., \& {Metzger}, B.~D. 2017, \mnras, 472, 3058, \dodoi{10.1093/mnras/stx2095}

\bibitem[{{Beniamini} {et~al.}(2019){Beniamini}, {Petropoulou}, {Barniol Duran}, \& {Giannios}}]{2019MNRAS.483..840B}
{Beniamini}, P., {Petropoulou}, M., {Barniol Duran}, R., \& {Giannios}, D. 2019, \mnras, 483, 840, \dodoi{10.1093/mnras/sty3093}

\bibitem[{{Bernuzzi} {et~al.}(2014){Bernuzzi}, {Dietrich}, {Tichy}, \& {Br{\"u}gmann}}]{Bernuzzi2014}
{Bernuzzi}, S., {Dietrich}, T., {Tichy}, W., \& {Br{\"u}gmann}, B. 2014, \prd, 89, 104021, \dodoi{10.1103/PhysRevD.89.104021}

\bibitem[{{Bernuzzi} {et~al.}(2020){Bernuzzi}, {Breschi}, {Daszuta}, {Endrizzi}, {Logoteta}, {Nedora}, {Perego}, {Radice}, {Schianchi}, {Zappa}, {Bombaci}, \& {Ortiz}}]{2020MNRAS.497.1488B}
{Bernuzzi}, S., {Breschi}, M., {Daszuta}, B., {et~al.} 2020, \mnras, 497, 1488, \dodoi{10.1093/mnras/staa1860}

\bibitem[{{Blandford} \& {Znajek}(1977)}]{1977MNRAS.179..433B}
{Blandford}, R.~D., \& {Znajek}, R.~L. 1977, \mnras, 179, 433, \dodoi{10.1093/mnras/179.3.433}

\bibitem[{{Bromberg} {et~al.}(2014){Bromberg}, {Granot}, {Lyubarsky}, \& {Piran}}]{2014MNRAS.443.1532B}
{Bromberg}, O., {Granot}, J., {Lyubarsky}, Y., \& {Piran}, T. 2014, \mnras, 443, 1532, \dodoi{10.1093/mnras/stu995}

\bibitem[{{Bromberg} {et~al.}(2015){Bromberg}, {Granot}, \& {Piran}}]{2015MNRAS.450.1077B}
{Bromberg}, O., {Granot}, J., \& {Piran}, T. 2015, \mnras, 450, 1077, \dodoi{10.1093/mnras/stv226}

\bibitem[{{Bromberg} {et~al.}(2012){Bromberg}, {Nakar}, {Piran}, \& {Sari}}]{2012ApJ...749..110B}
{Bromberg}, O., {Nakar}, E., {Piran}, T., \& {Sari}, R. 2012, \apj, 749, 110, \dodoi{10.1088/0004-637X/749/2/110}

\bibitem[{{Burns} {et~al.}(2023){Burns}, {Svinkin}, {Fenimore}, {Kann}, {Ag{\"u}{\'\i} Fern{\'a}ndez}, {Frederiks}, {Hamburg}, {Lesage}, {Temiraev}, {Tsvetkova}, {Bissaldi}, {Briggs}, {Dalessi}, {Dunwoody}, {Fletcher}, {Goldstein}, {Hui}, {Hristov}, {Kocevski}, {Lysenko}, {Mailyan}, {Mangan}, {McBreen}, {Racusin}, {Ridnaia}, {Roberts}, {Ulanov}, {Veres}, {Wilson-Hodge}, \& {Wood}}]{2023ApJ...946L..31B}
{Burns}, E., {Svinkin}, D., {Fenimore}, E., {et~al.} 2023, \apjl, 946, L31, \dodoi{10.3847/2041-8213/acc39c}

\bibitem[{{Camisasca} {et~al.}(2023){Camisasca}, {Guidorzi}, {Amati}, {Frontera}, {Song}, {Xiao}, {Xiong}, {Zhang}, {Margutti}, {Kobayashi}, {Mundell}, {Ge}, {Gomboc}, {Jia}, {Jordana-Mitjans}, {Li}, {Li}, {Maccary}, {Shrestha}, {Xue}, \& {Zhang}}]{2023A&A...671A.112C}
{Camisasca}, A.~E., {Guidorzi}, C., {Amati}, L., {et~al.} 2023, \aap, 671, A112, \dodoi{10.1051/0004-6361/202245657}

\bibitem[{{Cano} {et~al.}(2017){Cano}, {Wang}, {Dai}, \& {Wu}}]{2017AdAst2017E...5C}
{Cano}, Z., {Wang}, S.-Q., {Dai}, Z.-G., \& {Wu}, X.-F. 2017, Advances in Astronomy, 2017, 8929054, \dodoi{10.1155/2017/8929054}

\bibitem[{{Cowperthwaite} {et~al.}(2017){Cowperthwaite}, {Berger}, {Villar}, {Metzger}, {Nicholl}, {Chornock}, {Blanchard}, {Fong}, {Margutti}, {Soares-Santos}, {Alexander}, {Allam}, {Annis}, {Brout}, {Brown}, {Butler}, {Chen}, {Diehl}, {Doctor}, {Drout}, {Eftekhari}, {Farr}, {Finley}, {Foley}, {Frieman}, {Fryer}, {Garc{\'\i}a-Bellido}, {Gill}, {Guillochon}, {Herner}, {Holz}, {Kasen}, {Kessler}, {Marriner}, {Matheson}, {Neilsen}, {Quataert}, {Palmese}, {Rest}, {Sako}, {Scolnic}, {Smith}, {Tucker}, {Williams}, {Balbinot}, {Carlin}, {Cook}, {Durret}, {Li}, {Lopes}, {Louren{\c{c}}o}, {Marshall}, {Medina}, {Muir}, {Mu{\~n}oz}, {Sauseda}, {Schlegel}, {Secco}, {Vivas}, {Wester}, {Zenteno}, {Zhang}, {Abbott}, {Banerji}, {Bechtol}, {Benoit-L{\'e}vy}, {Bertin}, {Buckley-Geer}, {Burke}, {Capozzi}, {Carnero Rosell}, {Carrasco Kind}, {Castander}, {Crocce}, {Cunha}, {D'Andrea}, {da Costa}, {Davis}, {DePoy}, {Desai}, {Dietrich}, {Drlica-Wagner}, {Eifler}, {Evrard}, {Fernandez}, {Flaugher}, {Fosalba}, {Gaztanaga}, {Gerdes},
  {Giannantonio}, {Goldstein}, {Gruen}, {Gruendl}, {Gutierrez}, {Honscheid}, {Jain}, {James}, {Jeltema}, {Johnson}, {Johnson}, {Kent}, {Krause}, {Kron}, {Kuehn}, {Nuropatkin}, {Lahav}, {Lima}, {Lin}, {Maia}, {March}, {Martini}, {McMahon}, {Menanteau}, {Miller}, {Miquel}, {Mohr}, {Neilsen}, {Nichol}, {Ogando}, {Plazas}, {Roe}, {Romer}, {Roodman}, {Rykoff}, {Sanchez}, {Scarpine}, {Schindler}, {Schubnell}, {Sevilla-Noarbe}, {Smith}, {Smith}, {Sobreira}, {Suchyta}, {Swanson}, {Tarle}, {Thomas}, {Thomas}, {Troxel}, {Vikram}, {Walker}, {Wechsler}, {Weller}, {Yanny}, \& {Zuntz}}]{2017ApJ...848L..17C}
{Cowperthwaite}, P.~S., {Berger}, E., {Villar}, V.~A., {et~al.} 2017, \apjl, 848, L17, \dodoi{10.3847/2041-8213/aa8fc7}

\bibitem[{{Dainotti} {et~al.}(2022){Dainotti}, {De Simone}, {Islam}, {Kawaguchi}, {Moriya}, {Takiwaki}, {Tominaga}, \& {Gangopadhyay}}]{2022ApJ...938...41D}
{Dainotti}, M.~G., {De Simone}, B., {Islam}, K.~M., {et~al.} 2022, \apj, 938, 41, \dodoi{10.3847/1538-4357/ac8b77}

\bibitem[{{Eichler} {et~al.}(1989){Eichler}, {Livio}, {Piran}, \& {Schramm}}]{1989Natur.340..126E}
{Eichler}, D., {Livio}, M., {Piran}, T., \& {Schramm}, D.~N. 1989, \nat, 340, 126, \dodoi{10.1038/340126a0}

\bibitem[{{Fuller} \& {Ma}(2019)}]{2019ApJ...881L...1F}
{Fuller}, J., \& {Ma}, L. 2019, \apjl, 881, L1, \dodoi{10.3847/2041-8213/ab339b}

\bibitem[{{Ghirlanda} {et~al.}(2009){Ghirlanda}, {Nava}, {Ghisellini}, {Celotti}, \& {Firmani}}]{2009A&A...496..585G}
{Ghirlanda}, G., {Nava}, L., {Ghisellini}, G., {Celotti}, A., \& {Firmani}, C. 2009, \aap, 496, 585, \dodoi{10.1051/0004-6361/200811209}

\bibitem[{{Gottlieb} {et~al.}(2023){Gottlieb}, {Jacquemin-Ide}, {Lowell}, {Tchekhovskoy}, \& {Ramirez-Ruiz}}]{2023ApJ...952L..32G}
{Gottlieb}, O., {Jacquemin-Ide}, J., {Lowell}, B., {Tchekhovskoy}, A., \& {Ramirez-Ruiz}, E. 2023, \apjl, 952, L32, \dodoi{10.3847/2041-8213/ace779}

\bibitem[{{Gottlieb} {et~al.}(2024){Gottlieb}, {Renzo}, {Metzger}, {Goldberg}, \& {Cantiello}}]{2024arXiv240716745G}
{Gottlieb}, O., {Renzo}, M., {Metzger}, B.~D., {Goldberg}, J.~A., \& {Cantiello}, M. 2024, arXiv e-prints, arXiv:2407.16745, \dodoi{10.48550/arXiv.2407.16745}

\bibitem[{{Ivezi{\'c}} {et~al.}(2019){Ivezi{\'c}}, {Kahn}, {Tyson}, {Abel}, {Acosta}, {Allsman}, {Alonso}, {AlSayyad}, {Anderson}, {Andrew}, {Angel}, {Angeli}, {Ansari}, {Antilogus}, {Araujo}, {Armstrong}, {Arndt}, {Astier}, {Aubourg}, {Auza}, {Axelrod}, {Bard}, {Barr}, {Barrau}, {Bartlett}, {Bauer}, {Bauman}, {Baumont}, {Bechtol}, {Bechtol}, {Becker}, {Becla}, {Beldica}, {Bellavia}, {Bianco}, {Biswas}, {Blanc}, {Blazek}, {Blandford}, {Bloom}, {Bogart}, {Bond}, {Booth}, {Borgland}, {Borne}, {Bosch}, {Boutigny}, {Brackett}, {Bradshaw}, {Brandt}, {Brown}, {Bullock}, {Burchat}, {Burke}, {Cagnoli}, {Calabrese}, {Callahan}, {Callen}, {Carlin}, {Carlson}, {Chandrasekharan}, {Charles-Emerson}, {Chesley}, {Cheu}, {Chiang}, {Chiang}, {Chirino}, {Chow}, {Ciardi}, {Claver}, {Cohen-Tanugi}, {Cockrum}, {Coles}, {Connolly}, {Cook}, {Cooray}, {Covey}, {Cribbs}, {Cui}, {Cutri}, {Daly}, {Daniel}, {Daruich}, {Daubard}, {Daues}, {Dawson}, {Delgado}, {Dellapenna}, {de Peyster}, {de Val-Borro}, {Digel}, {Doherty}, {Dubois},
  {Dubois-Felsmann}, {Durech}, {Economou}, {Eifler}, {Eracleous}, {Emmons}, {Fausti Neto}, {Ferguson}, {Figueroa}, {Fisher-Levine}, {Focke}, {Foss}, {Frank}, {Freemon}, {Gangler}, {Gawiser}, {Geary}, {Gee}, {Geha}, {Gessner}, {Gibson}, {Gilmore}, {Glanzman}, {Glick}, {Goldina}, {Goldstein}, {Goodenow}, {Graham}, {Gressler}, {Gris}, {Guy}, {Guyonnet}, {Haller}, {Harris}, {Hascall}, {Haupt}, {Hernandez}, {Herrmann}, {Hileman}, {Hoblitt}, {Hodgson}, {Hogan}, {Howard}, {Huang}, {Huffer}, {Ingraham}, {Innes}, {Jacoby}, {Jain}, {Jammes}, {Jee}, {Jenness}, {Jernigan}, {Jevremovi{\'c}}, {Johns}, {Johnson}, {Johnson}, {Jones}, {Juramy-Gilles}, {Juri{\'c}}, {Kalirai}, {Kallivayalil}, {Kalmbach}, {Kantor}, {Karst}, {Kasliwal}, {Kelly}, {Kessler}, {Kinnison}, {Kirkby}, {Knox}, {Kotov}, {Krabbendam}, {Krughoff}, {Kub{\'a}nek}, {Kuczewski}, {Kulkarni}, {Ku}, {Kurita}, {Lage}, {Lambert}, {Lange}, {Langton}, {Le Guillou}, {Levine}, {Liang}, {Lim}, {Lintott}, {Long}, {Lopez}, {Lotz}, {Lupton}, {Lust}, {MacArthur}, {Mahabal},
  {Mandelbaum}, {Markiewicz}, {Marsh}, {Marshall}, {Marshall}, {May}, {McKercher}, {McQueen}, {Meyers}, {Migliore}, {Miller}, {Mills}, {Miraval}, {Moeyens}, {Moolekamp}, {Monet}, {Moniez}, {Monkewitz}, {Montgomery}, {Morrison}, {Mueller}, {Muller}, {Mu{\~n}oz Arancibia}, {Neill}, {Newbry}, {Nief}, {Nomerotski}, {Nordby}, {O'Connor}, {Oliver}, {Olivier}, {Olsen}, {O'Mullane}, {Ortiz}, {Osier}, {Owen}, {Pain}, {Palecek}, {Parejko}, {Parsons}, {Pease}, {Peterson}, {Peterson}, {Petravick}, {Libby Petrick}, {Petry}, {Pierfederici}, {Pietrowicz}, {Pike}, {Pinto}, {Plante}, {Plate}, {Plutchak}, {Price}, {Prouza}, {Radeka}, {Rajagopal}, {Rasmussen}, {Regnault}, {Reil}, {Reiss}, {Reuter}, {Ridgway}, {Riot}, {Ritz}, {Robinson}, {Roby}, {Roodman}, {Rosing}, {Roucelle}, {Rumore}, {Russo}, {Saha}, {Sassolas}, {Schalk}, {Schellart}, {Schindler}, {Schmidt}, {Schneider}, {Schneider}, {Schoening}, {Schumacher}, {Schwamb}, {Sebag}, {Selvy}, {Sembroski}, {Seppala}, {Serio}, {Serrano}, {Shaw}, {Shipsey}, {Sick}, {Silvestri},
  {Slater}, {Smith}, {Smith}, {Sobhani}, {Soldahl}, {Storrie-Lombardi}, {Stover}, {Strauss}, {Street}, {Stubbs}, {Sullivan}, {Sweeney}, {Swinbank}, {Szalay}, {Takacs}, {Tether}, {Thaler}, {Thayer}, {Thomas}, {Thornton}, {Thukral}, {Tice}, {Trilling}, {Turri}, {Van Berg}, {Vanden Berk}, {Vetter}, {Virieux}, {Vucina}, {Wahl}, {Walkowicz}, {Walsh}, {Walter}, {Wang}, {Wang}, {Warner}, {Wiecha}, {Willman}, {Winters}, {Wittman}, {Wolff}, {Wood-Vasey}, {Wu}, {Xin}, {Yoachim}, \& {Zhan}}]{2019ApJ...873..111I}
{Ivezi{\'c}}, {\v{Z}}., {Kahn}, S.~M., {Tyson}, J.~A., {et~al.} 2019, \apj, 873, 111, \dodoi{10.3847/1538-4357/ab042c}

\bibitem[{{Jacquemin-Ide} {et~al.}(2024{\natexlab{a}}){Jacquemin-Ide}, {Gottlieb}, {Lowell}, \& {Tchekhovskoy}}]{2024ApJ...961..212J}
{Jacquemin-Ide}, J., {Gottlieb}, O., {Lowell}, B., \& {Tchekhovskoy}, A. 2024{\natexlab{a}}, \apj, 961, 212, \dodoi{10.3847/1538-4357/ad02f0}

\bibitem[{{Jacquemin-Ide} {et~al.}(2024{\natexlab{b}}){Jacquemin-Ide}, {Rincon}, {Tchekhovskoy}, \& {Liska}}]{2024MNRAS.532.1522J}
{Jacquemin-Ide}, J., {Rincon}, F., {Tchekhovskoy}, A., \& {Liska}, M. 2024{\natexlab{b}}, \mnras, 532, 1522, \dodoi{10.1093/mnras/stae1538}

\bibitem[{{James} {et~al.}(2022){James}, {Janiuk}, \& {Nouri}}]{2022ApJ...935..176J}
{James}, B., {Janiuk}, A., \& {Nouri}, F.~H. 2022, \apj, 935, 176, \dodoi{10.3847/1538-4357/ac81b7}

\bibitem[{{Janiuk} \& {James}(2022)}]{2022A&A...668A..66J}
{Janiuk}, A., \& {James}, B. 2022, \aap, 668, A66, \dodoi{10.1051/0004-6361/202244196}

\bibitem[{{Kann} {et~al.}(2023){Kann}, {Agayeva}, {Aivazyan}, {Alishov}, {Andrade}, {Antier}, {Baransky}, {Bendjoya}, {Benkhaldoun}, {Beradze}, {Berezin}, {Bo{\"e}r}, {Broens}, {Brunier}, {Bulla}, {Burkhonov}, {Burns}, {Chen}, {Chen}, {Conti}, {Coughlin}, {Cui}, {Daigne}, {Delaveau}, {Devillepoix}, {Dietrich}, {Dornic}, {Dubois}, {Ducoin}, {Durand}, {Duverne}, {Eggenstein}, {Ehgamberdiev}, {Fouad}, {Freeberg}, {Froebrich}, {Ge}, {Gervasoni}, {Godunova}, {Gokuldass}, {Gurbanov}, {Han}, {Hasanov}, {Hello}, {Hussenot-Desenonges}, {Inasaridze}, {Iskandar}, {Ismailov}, {Janati}, {du Laz}, {Jia}, {Karpov}, {Kaeouach}, {Kiendrebeogo}, {Klotz}, {Kneip}, {Kochiashvili}, {Kunert}, {Lekic}, {Leonini}, {Li}, {Li}, {Li}, {Liao}, {Logie}, {Lu}, {Mao}, {Marchais}, {M{\'e}nard}, {Morris}, {Natsvlishvili}, {Nedora}, {Noonan}, {Noysena}, {Orange}, {Pang}, {Peng}, {Pellouin}, {Peloton}, {Pradier}, {Pyshna}, {Rajabov}, {Rau}, {Rinner}, {Rivet}, {Romanov}, {Rosi}, {Rupchandani}, {Serrau}, {Shokry}, {Simon}, {Smith}, {Sokoliuk},
  {Soliman}, {Song}, {Takey}, {Tillayev}, {Ramirez}, {e Melo}, {Turpin}, {de Ugarte Postigo}, {Vanaverbeke}, {Vasylenko}, {Vernet}, {Vidadi}, {Wang}, {Wang}, {Wang}, {Wang}, {Xiong}, {Xu}, {Xue}, {Zeng}, {Zhang}, {Zhao}, \& {Zhao}}]{2023ApJ...948L..12K}
{Kann}, D.~A., {Agayeva}, S., {Aivazyan}, V., {et~al.} 2023, \apjl, 948, L12, \dodoi{10.3847/2041-8213/acc8d0}

\bibitem[{{Kastaun} {et~al.}(2013){Kastaun}, {Galeazzi}, {Alic}, {Rezzolla}, \& {Font}}]{Kastaun2013}
{Kastaun}, W., {Galeazzi}, F., {Alic}, D., {Rezzolla}, L., \& {Font}, J.~A. 2013, \prd, 88, 021501, \dodoi{10.1103/PhysRevD.88.021501}

\bibitem[{{Kiuchi} {et~al.}(2014){Kiuchi}, {Kyutoku}, {Sekiguchi}, {Shibata}, \& {Wada}}]{2014PhRvD..90d1502K}
{Kiuchi}, K., {Kyutoku}, K., {Sekiguchi}, Y., {Shibata}, M., \& {Wada}, T. 2014, \prd, 90, 041502, \dodoi{10.1103/PhysRevD.90.041502}

\bibitem[{{Leng} \& {Giannios}(2014)}]{2014MNRAS.445L...1L}
{Leng}, M., \& {Giannios}, D. 2014, \mnras, 445, L1, \dodoi{10.1093/mnrasl/slu122}

\bibitem[{{Lesage} {et~al.}(2023){Lesage}, {Veres}, {Briggs}, {Goldstein}, {Kocevski}, {Burns}, {Wilson-Hodge}, {Bhat}, {Huppenkothen}, {Fryer}, {Hamburg}, {Racusin}, {Bissaldi}, {Cleveland}, {Dalessi}, {Fletcher}, {Giles}, {Hristov}, {Hui}, {Mailyan}, {Malacaria}, {Poolakkil}, {Roberts}, {von Kienlin}, {Wood}, {Ajello}, {Arimoto}, {Baldini}, {Ballet}, {Baring}, {Bastieri}, {Gonzalez}, {Bellazzini}, {Bissaldi}, {Blandford}, {Bonino}, {Bruel}, {Buson}, {Cameron}, {Caputo}, {Caraveo}, {Cavazzuti}, {Chiaro}, {Cibrario}, {Ciprini}, {Orestano}, {Crnogorcevic}, {Cuoco}, {Cutini}, {D'Ammando}, {De Gaetano}, {Di Lalla}, {Di Venere}, {Dom{\'\i}nguez}, {Fegan}, {Ferrara}, {Fleischhack}, {Fukazawa}, {Funk}, {Fusco}, {Galanti}, {Gammaldi}, {Gargano}, {Gasbarra}, {Gasparrini}, {Germani}, {Giacchino}, {Giglietto}, {Gill}, {Giroletti}, {Granot}, {Green}, {Grenier}, {Guiriec}, {Gustafsson}, {Hays}, {Hewitt}, {Horan}, {Hou}, {Kuss}, {Latronico}, {Laviron}, {Lemoine-Goumard}, {Li}, {Liodakis}, {Longo}, {Loparco}, {Lorusso},
  {Lovellette}, {Lubrano}, {Maldera}, {Manfreda}, {Mart{\'\i}-Devesa}, {Mazziotta}, {McEnery}, {Mereu}, {Meyer}, {Michelson}, {Mizuno}, {Monzani}, {Morselli}, {Moskalenko}, {Negro}, {Nuss}, {Omodei}, {Orlando}, {Ormes}, {Paneque}, {Panzarini}, {Persic}, {Pesce-Rollins}, {Pillera}, {Piron}, {Poon}, {Porter}, {Principe}, {Rain{\`o}}, {Rando}, {Rani}, {Razzano}, {Razzaque}, {Reimer}, {Reimer}, {Ryde}, {S{\'a}nchez-Conde}, {Parkinson}, {Scotton}, {Serini}, {Sgr{\`o}}, {Sharma}, {Siskind}, {Spandre}, {Spinelli}, {Tajima}, {Torres}, {Valverde}, {Venters}, {Wadiasingh}, {Wood}, \& {Zaharijas}}]{2023ApJ...952L..42L}
{Lesage}, S., {Veres}, P., {Briggs}, M.~S., {et~al.} 2023, \apjl, 952, L42, \dodoi{10.3847/2041-8213/ace5b4}

\bibitem[{{Liu} {et~al.}(2015){Liu}, {Hou}, {Xue}, \& {Gu}}]{2015ApJS..218...12L}
{Liu}, T., {Hou}, S.-J., {Xue}, L., \& {Gu}, W.-M. 2015, \apjs, 218, 12, \dodoi{10.1088/0067-0049/218/1/12}

\bibitem[{{Lowell} {et~al.}(2024){Lowell}, {Jacquemin-Ide}, {Tchekhovskoy}, \& {Duncan}}]{2024ApJ...960...82L}
{Lowell}, B., {Jacquemin-Ide}, J., {Tchekhovskoy}, A., \& {Duncan}, A. 2024, \apj, 960, 82, \dodoi{10.3847/1538-4357/ad09af}

\bibitem[{{Lyutikov} \& {Blandford}(2003)}]{2003astro.ph.12347L}
{Lyutikov}, M., \& {Blandford}, R. 2003, arXiv e-prints, astro, \dodoi{10.48550/arXiv.astro-ph/0312347}

\bibitem[{{Mazzali} {et~al.}(2014){Mazzali}, {McFadyen}, {Woosley}, {Pian}, \& {Tanaka}}]{2014MNRAS.443...67M}
{Mazzali}, P.~A., {McFadyen}, A.~I., {Woosley}, S.~E., {Pian}, E., \& {Tanaka}, M. 2014, \mnras, 443, 67, \dodoi{10.1093/mnras/stu1124}

\bibitem[{{McBreen} {et~al.}(2002){McBreen}, {McBreen}, {Hanlon}, \& {Quilligan}}]{2002A&A...393L..29M}
{McBreen}, S., {McBreen}, B., {Hanlon}, L., \& {Quilligan}, F. 2002, \aap, 393, L29, \dodoi{10.1051/0004-6361:20021073}

\bibitem[{{McKinney} {et~al.}(2012){McKinney}, {Tchekhovskoy}, \& {Blandford}}]{2012MNRAS.423.3083M}
{McKinney}, J.~C., {Tchekhovskoy}, A., \& {Blandford}, R.~D. 2012, \mnras, 423, 3083, \dodoi{10.1111/j.1365-2966.2012.21074.x}

\bibitem[{{Metzger} {et~al.}(2018){Metzger}, {Beniamini}, \& {Giannios}}]{2018ApJ...857...95M}
{Metzger}, B.~D., {Beniamini}, P., \& {Giannios}, D. 2018, \apj, 857, 95, \dodoi{10.3847/1538-4357/aab70c}

\bibitem[{{Metzger} {et~al.}(2011){Metzger}, {Giannios}, {Thompson}, {Bucciantini}, \& {Quataert}}]{2011MNRAS.413.2031M}
{Metzger}, B.~D., {Giannios}, D., {Thompson}, T.~A., {Bucciantini}, N., \& {Quataert}, E. 2011, \mnras, 413, 2031, \dodoi{10.1111/j.1365-2966.2011.18280.x}

\bibitem[{{Moderski} \& {Sikora}(1996)}]{1996MNRAS.283..854M}
{Moderski}, R., \& {Sikora}, M. 1996, \mnras, 283, 854, \dodoi{10.1093/mnras/283.3.854}

\bibitem[{{Novikov} \& {Thorne}(1973)}]{1973blho.conf..343N}
{Novikov}, I.~D., \& {Thorne}, K.~S. 1973, in Black Holes (Les Astres Occlus), ed. C.~{Dewitt} \& B.~S. {Dewitt}, 343--450

\bibitem[{{O'Connor} {et~al.}(2023){O'Connor}, {Troja}, {Ryan}, {Beniamini}, {van Eerten}, {Granot}, {Dichiara}, {Ricci}, {Lipunov}, {Gillanders}, {Gill}, {Moss}, {Anand}, {Andreoni}, {Becerra}, {Buckley}, {Butler}, {Cenko}, {Chasovnikov}, {Durbak}, {Francile}, {Hammerstein}, {van der Horst}, {Kasliwal}, {Kouveliotou}, {Kutyrev}, {Lee}, {Srinivasaragavan}, {Topolev}, {Watson}, {Yang}, \& {Zhirkov}}]{2023SciA....9I1405O}
{O'Connor}, B., {Troja}, E., {Ryan}, G., {et~al.} 2023, Science Advances, 9, eadi1405, \dodoi{10.1126/sciadv.adi1405}

\bibitem[{{Rouco Escorial} {et~al.}(2023){Rouco Escorial}, {Fong}, {Berger}, {Laskar}, {Margutti}, {Schroeder}, {Rastinejad}, {Cornish}, {Popp}, {Lally}, {Nugent}, {Paterson}, {Metzger}, {Chornock}, {Alexander}, {Cendes}, \& {Eftekhari}}]{2023ApJ...959...13R}
{Rouco Escorial}, A., {Fong}, W., {Berger}, E., {et~al.} 2023, \apj, 959, 13, \dodoi{10.3847/1538-4357/acf830}

\bibitem[{{Sander} {et~al.}(2019){Sander}, {Hamann}, {Todt}, {Hainich}, {Shenar}, {Ramachandran}, \& {Oskinova}}]{2019A&A...621A..92S}
{Sander}, A.~A.~C., {Hamann}, W.~R., {Todt}, H., {et~al.} 2019, \aap, 621, A92, \dodoi{10.1051/0004-6361/201833712}

\bibitem[{{Shahmoradi} \& {Nemiroff}(2015)}]{2015MNRAS.451..126S}
{Shahmoradi}, A., \& {Nemiroff}, R.~J. 2015, \mnras, 451, 126, \dodoi{10.1093/mnras/stv714}

\bibitem[{{Spruit}(2002)}]{2002A&A...381..923S}
{Spruit}, H.~C. 2002, \aap, 381, 923, \dodoi{10.1051/0004-6361:20011465}

\bibitem[{{Stanek} {et~al.}(2003){Stanek}, {Matheson}, {Garnavich}, {Martini}, {Berlind}, {Caldwell}, {Challis}, {Brown}, {Schild}, {Krisciunas}, {Calkins}, {Lee}, {Hathi}, {Jansen}, {Windhorst}, {Echevarria}, {Eisenstein}, {Pindor}, {Olszewski}, {Harding}, {Holland}, \& {Bersier}}]{2003ApJ...591L..17S}
{Stanek}, K.~Z., {Matheson}, T., {Garnavich}, P.~M., {et~al.} 2003, \apjl, 591, L17, \dodoi{10.1086/376976}

\bibitem[{{Tchekhovskoy} \& {Giannios}(2015)}]{2015MNRAS.447..327T}
{Tchekhovskoy}, A., \& {Giannios}, D. 2015, \mnras, 447, 327, \dodoi{10.1093/mnras/stu2229}

\bibitem[{{Tchekhovskoy} {et~al.}(2010){Tchekhovskoy}, {Narayan}, \& {McKinney}}]{2010ApJ...711...50T}
{Tchekhovskoy}, A., {Narayan}, R., \& {McKinney}, J.~C. 2010, \apj, 711, 50, \dodoi{10.1088/0004-637X/711/1/50}

\bibitem[{{Tchekhovskoy} {et~al.}(2011){Tchekhovskoy}, {Narayan}, \& {McKinney}}]{2011MNRAS.418L..79T}
---. 2011, \mnras, 418, L79, \dodoi{10.1111/j.1745-3933.2011.01147.x}

\bibitem[{{Usov}(1992)}]{1992Natur.357..472U}
{Usov}, V.~V. 1992, \nat, 357, 472, \dodoi{10.1038/357472a0}

\bibitem[{{Wanderman} \& {Piran}(2010)}]{2010MNRAS.406.1944W}
{Wanderman}, D., \& {Piran}, T. 2010, \mnras, 406, 1944, \dodoi{10.1111/j.1365-2966.2010.16787.x}

\bibitem[{{Wang} {et~al.}(2018){Wang}, {Zhang}, {Liang}, {Lu}, {Lin}, {Li}, \& {Li}}]{2018ApJ...859..160W}
{Wang}, X.-G., {Zhang}, B., {Liang}, E.-W., {et~al.} 2018, \apj, 859, 160, \dodoi{10.3847/1538-4357/aabc13}

\bibitem[{{Woosley}(1993)}]{1993ApJ...405..273W}
{Woosley}, S.~E. 1993, \apj, 405, 273, \dodoi{10.1086/172359}

\bibitem[{{Woosley} \& {Bloom}(2006)}]{2006ARA&A..44..507W}
{Woosley}, S.~E., \& {Bloom}, J.~S. 2006, \araa, 44, 507, \dodoi{10.1146/annurev.astro.43.072103.150558}

\bibitem[{{Zalamea} \& {Beloborodov}(2011)}]{2011MNRAS.410.2302Z}
{Zalamea}, I., \& {Beloborodov}, A.~M. 2011, \mnras, 410, 2302, \dodoi{10.1111/j.1365-2966.2010.17600.x}

\bibitem[{{Zitouni} {et~al.}(2018){Zitouni}, {Guessoum}, {AlQassimi}, \& {Alaryani}}]{2018Ap&SS.363..223Z}
{Zitouni}, H., {Guessoum}, N., {AlQassimi}, K.~M., \& {Alaryani}, O. 2018, \apss, 363, 223, \dodoi{10.1007/s10509-018-3449-0}

\end{thebibliography}
\bibliographystyle{aasjournal}

\end{document}